# Discharge structure hierarchy of highly electronegative plasma and the implication on nuclear fusion at low pressure and quasi- cold ions approximation

Yuan-He Sun[1], Shu-Xia Zhao[1]*, Rui-Ji Tang[1], Yu Tian[2]

[1]School of Physics, Dalian University of Technology, Dalian, 116024, China

[2]Experimental training center, Dalian University of Science and Technology, Dalian, 116052, China

Correspondence: zhaonie@dlut.edu.cn

**Abstract**

In this paper, the discharge structure of an Ar/SF$_6$ inductively coupled plasma (ICP) at the low pressure, 10 mTorr, is investigated by the fluid simulation at the quasi- cold ions approximation, i.e., room temperature. The structure is found to be hierarchal and in this simulated hierarchy, the stratification, the parabola profile in the stratified core, the double layer, and the coagulated profile in the core center are examined. This fluid simulation version and a quasi- fluid simulation of an Ar/CF$_4$ ICP given by the HPEM code, cooperatively enlighten the discharge structure of highly electronegative ICPs and meanwhile suggest the potential applications of them. It is found that when the ions are cold the hierarchy is predicted and when the ions are thermalized the simple discharge structure appears. In the simulated hierarchy, the double layer formed at the interface of halo and core is given by the ionic and acoustic vibrations. The simulated cations flux and potential at the two sides of the double layer are double-valued and the simulated double layer is modelled as the dipole microscopically and as the capacitor macroscopically. The evolution of discharge structure hierarchy is presented, in which the harmony among many processes in the hierarchy are achieved and the charm of self-coagulation is exhibited, i.e., the bigger is the coagulated volume, the denser is the coagulated mass. This provides insights for creating possibly the free nuclear fusion by means of self-coagulation, and this type of nuclear fusion without the inertial and magnetic constrictions is achieved by means of the compressible heating scheme. The self-coagulation helps people re-recognize the anions Boltzmann balance, and meanwhile it turns the discharging and collisional plasmas into the collisionless and astrophysical plasmas, in which the double layer and ionization instability occur.

## I. Introduction

The low-pressure and radio-frequency (RF) electronegative plasma sources are widely used in the Si-based material etching processes and the functional thin-film deposition [1, 2]. In general, the capacitive [3] and inductive radio frequency sources are applied, and the inductively coupled plasma [4, 5] (abbreviated as the ICP) is characterized by its low-pressure discharge, high plasma density, and simple device. At present, the discharge structure of the plasma source characterizes the profiles of plasma species densities, and the related theories illustrate the underlying physics and chemistry that give rise to the profiles. The physical process involves ambipolar diffusion, and the chemistry involves inelastic collisions of electron impact and reactions of heavy species.

Moreover, the plasma collective interaction, the Boltzmann balance, and the Bohm sheath criterion play important roles in forming the steady-state density profiles. Owing to the presence of anions in the discharge, the physical and chemical processes occurring in the electronegative plasmas are complex, and their discharge structures are hierarchical compared to that of electropositive plasmas. Studies of the discharge structure are important since they help people understand the mechanisms that occur in the discharging plasmas, and more importantly are helpful for exciting new ideas on possible novel applications of them, e.g., the free nuclear fusion, based on the deep recognitions on the electronegative plasma properties, e.g., the self-coagulation.

In the last century, the discharge structure of electronegative plasma has been extensively investigated, based on the theoretical analysis and simple numerical calculation of fluid model. In Refs. [6-9], the stratification of the electronegative discharge structure into an electronegative core consisting of cations and anions (with a small number of electrons) and an electropositive halo consisting of cations and electrons (with tiny anions) were investigated, at the approximation of cold ions, i.e., neglecting the diffusion of ions at the condition, $T_i = 0$. Herein, $T_i$ is the ions temperature. Furthermore, in Ref. [10] the effect of finite ion temperature was included, and it was found that the central anions and cations core is preserved but the previously shape boundary is progressively smoothed. Besides, in Ref. [11] the Boltzmann balance was used for both the anions and electrons, and the parabola profile in the electronegative core was given by the theoretical analysis, at further neglecting the bulk ionic recombination loss therein, which is only validated at the low pressure. There were then the intense arguments on whether the anions can be assumed to the Boltzmann balance or not in Refs. [12-14], and ultimately the arguments were unified by Bogdonov in Ref. [15] that the anions Boltzmann balance is only realized at the low pressure when the role of attachment is small as compared to the ambi-polar diffusion of anions, which was later validated by the experimental work in Ref. [16].

The above works were based on the electrically neutral plasma approximation, and when the Poisson equation was used the double layer was found at the interface of the electronegative core and the electropositive halo in Refs. [17-22]. Specially in Refs. [19, 20], in a low-pressure and collisionless electronegative discharge, for the plasma approximation, the plasma edge potential is double valued when the anions are sufficiently cold. When the strict charge neutrality is relaxed, the double layer was observed at the edge of plasma, at the condition that the flux associated with low (in absolute value) potential solution is less than that of the high potential solution. Herein, the plasma edge is meant the interface of core and halo. The collisionless fluid model is meant the famous Tonks-Langmuir model as illustrated in Refs. [23, 24]. Besides for the fluid model, the kinetic model was used for investigating the electronegative discharge structure as well in Ref. [21], aimed at relaxing the anions Boltzmann assumption. It was found in this reference, the double-layer stratified discharge is observed when the effective anions temperature is sufficiently low, in qualitative agreement with the fluid model prediction with the anions Boltzmann balance. Besides for the above collisionless fluid and kinetic models, the effects of ion-neutral collisions on the discharge structure, and the double layer and flux, were investigated in Ref. [22], and it was found that the typical three types of discharge, i.e., stratified one, double layer stratified one, and the uniform one, given by the collisionless models above, were predicted after including the collisions. In addition, the flux and potential at the collisional plasma edge have the same characteristics as in the collisionless case. It is noted the addition of collisions in the fluid model in Ref. [22] just adjusted the ambi-polar potential, through which the discharge structure is affected. The realistic meanings of collisional plasma, as seen in the Secs. (3.6-3.8) of present article, are the electric conductivity and the Ohm heating mechanism. Ultimately, the condition that the anions are sufficiently cold is stressed frequently above for realizing the double layer stratified structure. This fact is also validated by our fluid simulations with two types of code (see Secs. (2.1) and (3.7)) that the complex structure consisting of the stratification and double layer occurs when the ions are in the room temperature range, and the discharge hierarchy disappears when the ions are thermalized.

This room temperature condition of anions is thereby meant *quasi-* cold ions approximation, as distinguishing with the *concise* cold ions approximation, that is, zero temperature in Kevin unit as designed in the early analytical works of Refs. [6-9].

It is noted that the early works of electronegative discharge structure [6-25] are not self-consistent and their discoverings on the structure of plasma are thereby not complete. For instance, in Ref. [25], the energy deposition process of an ICP $SF_6$ discharge was assumed to be spatially uniform in a detailed fluid model and the simulated discharge structure was found to evolve from the parabola to the ellipse profile at increasing the pressure. Besides for the parabola at low pressure, the ellipse profile at the high pressure can also be predicted by the analytical theory, as described in Ref. [26]. It is stressed that although the basic evolution of discharge structure from the parabola to the ellipse is given by this detailed fluid model, the hierarchy and the self-coagulation dynamics are not displayed, due to the less self-consistent treatment of the ICP energy deposition process. Our presently self-consistent fluid simulation of an Ar/$SF_6$ ICP source at the low pressure, although with the quasi- cold ions approximation, are shown in the present article, i.e., the Secs. (3.1-3.4). It is found in these sections the discharge structure is hierarchal and it includes the stratification, the parabola profile in the core, the double layer and the self-coagulation, etc. The self-coagulation [27, 28] is the dominated factor in the electronegative discharge. Besides, an overview of the discharge structure hierarchy at the high pressure that includes the self-coagulation, the ellipse profile, and the ionization instability, is presented in the Secs. (3.5) and (3.6) of article for a good comparison to the case of low pressure. Furthermore, a quasi- fluid simulation of an Ar/$CF_4$ mixed ICP of us in which the ions are thermalized is shown in the Sec. (3.7) of article, and it shows that the discharge structure hierarchy is not occurred but a simple discharge structure appears at the thermal ions. It is then speculated when decreasing the gas flow rate in the thermalized ionic plasma, the simple discharge structure may transfer to the discharge structure hierarchy since the ions are possibly quenched at the attenuation of the Ohm heating scheme. This probably leads to the nuclear fusion [29, 30] that is occurred without the help of constrictions of inertial term and magnetic field, but relying on the self-coagulation and the compressible heating mechanism, as designed in the Sec. (3.8) of article. Besides, the self-coagulation plays the role of turning the collisional plasma to the collisionless plasma, in which the astrophysical plasma behaviors, e.g., the double layer [31-33] and the ionization instability [34-41], are occurred. Moreover, it provides new insights for re-recognizing the anions Boltzmann balance. Lastly, it is noted that all these findings related to the discharge structure of highly electronegative plasma and the physics of them as summarized above are given by the collaboration between the self-consistent simulations and the analytical theories, and hence the relation between them needs to be clarified. Namely, the theories provide the behind physics that helps people understand the profiles of quantities generated from the simulations, such as the parabolic and elliptic ionic density profiles, and meanwhile the complete simulations provide novel phenomena that excite new physical models and related theories, e.g., self-coagulation. They collaborate to assist people for better understanding complex systems, such as the electronegative discharging plasma reported herein.

The remainder of this article is organized as follows. In Sec. 2, firstly, the fluid model of the Ar/$SF_6$ ICP source at the quasi cold ions approximation is described. Then, the related discharge structure theories, parabola and self-coagulation, are presented. In Sec. 3 the results and discussion are given and in Sec. 4 the conclusion and further remarks are presented.

## II. Methodology

### (2.1) Fluid model of Ar/$SF_6$ ICP source

In this study, the fluid model is used to simulate the Ar/$SF_6$ mixed ICP source at the pressures, 10 and 90 mTorr, at the fixed Ar versus $SF_6$ ratio, 9:1, and the fixed power, 300W. The behind physics and chemistry of these simulations are partially explained based on the discharge structure theories given in the next Secs. (2.2) and (2.3). The equations of electron density and energy of fluid model are given as follows.

$$\frac{\partial n_e}{\partial t} + \nabla \cdot \mathbf{\Gamma}_e = R_e, \tag{1}$$

$$\frac{\partial n_\varepsilon}{\partial t} + \nabla \cdot \mathbf{\Gamma}_\varepsilon + \mathbf{E} \cdot \mathbf{\Gamma}_e = R_\varepsilon + P_{ohm}. \tag{2}$$

Herein, $n_e$ and $n_\varepsilon$ are the number density and energy density of electrons, respectively. $\mathbf{\Gamma}_e$ and $\mathbf{\Gamma}_\varepsilon$ are the corresponding fluxes of number density and energy density of electrons, respectively. $R_e$ and $R_\varepsilon$ are the corresponding source terms for the number density and energy density of electrons, respectively. $P_{ohm}$ is the deposited power density via the electron Ohm heating scheme of RF field in the azimuthal direction, i.e., $P_{ohm} = \frac{1}{2}\text{Re}(\sigma_e |E_\theta|^2)$, in which $\sigma_e$ is the electrical conductivity of electrons and $E_\theta$ is the azimuthal RF field (see next). It is seen that the energy deposition process of ICP source in our selected fluid model is self-consistent. And, $\mathbf{E}$ is the electrostatic field vector in the radial and axial directions.

The mass fraction equation is used to describe the mass transport of heavy species in Eq. (3).

$$\rho \frac{\partial w_k}{\partial t} = \nabla \cdot \mathbf{j}_k + R_k. \tag{3}$$

Herein, $\rho$ is the total mass density of heavy species and $w_k$ is the mass fraction of species $k$. $\mathbf{j}_k$ is the diffusive and drift flux of species $k$. It is noted the heavy species temperature is assumed to be room temperature, which satisfies the quasi- cold ions approximation as mentioned in Sec. I.

To describe the azimuthal RF field in the reactor, the Maxwell equations are combined to express the Ampere law in Eq. (4).

$$(j\omega\sigma_e - \omega^2 \varepsilon_0 \varepsilon_r)\mathbf{A} + \nabla \times (\mu_0^{-1}\mu_r^{-1}\nabla \times \mathbf{A}) = \mathbf{J}_a. \tag{4}$$

Herein, $j$ is the imaginary unit and $\omega$ is the angular frequency of power source, expressed as $2\pi f$ at $f = 13.56 \text{MHz}$. $\varepsilon_0$ and $\varepsilon_r$ are the vacuum permittivity and the relative permittivity of dielectric window material (quartz), respectively. $\mu_0$ and $\mu_r$ are the vacuum permeability and the relative permeability of coil that is made of copper, respectively. $\mathbf{A}$ is the magnetic vector potential, from which the RF magnetic and electric fields are calculated as, $\mathbf{B} = \nabla \times \mathbf{A}, \mathbf{E}_{rf} = -\frac{\partial \mathbf{A}}{\partial t}$. When considering the azimuthal symmetry, only the azimuthal component of RF electric field, $E_\theta$, and the axial and radial components of RF magnetic field, $B_r, B_z$, need to be addressed. $\mathbf{J}_a$ is the applied coil current density and its magnitude is varied until the required power, 300 W, is achieved.

The Poisson equation is used to calculate the electrostatic field in Eq. (5).

$$\begin{aligned}\mathbf{E} &= -\nabla V, \\ \nabla \cdot \mathbf{D} &= \rho_V.\end{aligned} \quad (5)$$

Herein, $\mathbf{D}$ is the electric displacement vector, from which the electrostatic field can be deduced through the permittivity, and $\rho_V$ is the spatial charge density.

The Ar/SF$_6$ gas-phase chemistry includes the elastic collision, the excitation and deexcitation, the ionization, the direct and dissociative attachments, and the dissociation of the electron-impact. The heavy species reactions include the neutral and ionic recombination, the detachment, the Penning ionization and the charge exchange. The surface kinetics of species include the recombination and the de-excitation. The ICP reactor consists of the vacuum chamber, the dielectric window and the plasma chamber. The radius of dielectric window and vacuum chamber is 14 cm, and the heights of them are 1 cm and 3 cm, respectively. The radius of plasma chamber is 15 cm and the height of it is 13 cm. To simulate the plasma processing technique, the substrate with a radius of 13 cm and a height of 4 cm is seated at the bottom center of plasma chamber. The two-turn coil is installed above the dielectric window and the radial locations of two turns are 8 cm and 10 cm, respectively. More detail about the fluid model of Ar/SF$_6$ ICP source can be found in Ref. [28].

**(2.2) Theory of core parabola**

The balances of flux and charge density in the electronegative plasma of core stratified are given in Eqs. (6, 7). Here, $\Gamma_+$, $\Gamma_-$, and $\Gamma_e$ are the fluxes of cations, anions, and electrons, respectively. $n_+$, $n_-$, and $n_e$ are the densities of the cations, anions, and electrons, respectively. The electronegativity $\alpha$ is defined in Eq. (8), i.e., the ratio between the anions density and electrons density. The fluxes of cations, anions, and electrons at the drift and diffusion approximations are given in Eqs. (9-11). Herein, $D_+$, $D_-$, and $D_e$ are the diffusion coefficients of the cations, anions, and electrons, respectively. $\mu_+$, $\mu_-$, and $\mu_e$ are the mobilities of the cations, anions, and electrons, respectively. At correlating Eqs. (6-11), the flux of cations is rewritten as a function of the quantities, $\mu_+, \mu_-, \mu_e$, $D_+, D_-, D_e$, $n_+, n_-, n_e$ and $\alpha$, in Eq. (12). Like the electropositive plasma, the ambipolar diffusion coefficient of electronegative plasma, $D_{a+}$, is introduced in Eq. (13). The cations flux is then re-expressed as the product of the newly introduced coefficient and the cations density gradient, as illustrated in Eq. (14).

$$\Gamma_+ = \Gamma_- + \Gamma_e, \tag{6}$$

$$n_+ = n_- + n_e, \tag{7}$$

$$\alpha = n_- / n_e, \tag{8}$$

$$\Gamma_+ = -D_+ \nabla n_+ + n_+ \mu_+ \mathbf{E}, \tag{9}$$

$$\Gamma_- = -D_- \nabla n_- - n_- \mu_- \mathbf{E}, \tag{10}$$

$$\Gamma_e = -D_e \nabla n_e - n_e \mu_e \mathbf{E}, \tag{11}$$

$$\Gamma_+ = -\frac{(\mu_e + \mu_- \alpha) D_+ + \mu_+ (1+\alpha) D_e (\nabla n_e / \nabla n_+) + \mu_+ (1+\alpha) D_- (\nabla n_- / \nabla n_+)}{\mu_e + \mu_- \alpha + \mu_+ (1+\alpha)} \nabla n_+, \tag{12}$$

$$D_{a+} = \frac{(\mu_e + \mu_- \alpha) D_+ + \mu_+ (1+\alpha) D_e (\nabla n_e / \nabla n_+) + \mu_+ (1+\alpha) D_- (\nabla n_- / \nabla n_+)}{\mu_e + \mu_- \alpha + \mu_+ (1+\alpha)}, \tag{13}$$

$$\Gamma_+ = -D_{a+} \nabla n_+. \tag{14}$$

As seen above, this version of ambipolar diffusion coefficient in Eq. (13) is complicated. Therefore, reasonable approximations are need to reduce its complexity and then clear physics can emerge. To achieve this goal, the huge difference between the electron and anion temperatures that characterizes the non- thermal equilibrium plasma is utilized. In Eq. (15), the parameter $\gamma$, is introduced, which is defined as the ratio between the electron and anion temperatures. Here, $T_e$ and $T_i$ are the electron and anion temperatures, respectively. It is noticed that the anion and cation are in thermal equilibrium; hence, their temperatures are the same. The Boltzmann balance is then adopted for both the electrons and anions, and the relation between the electron and anion densities is found, as shown in Eq. (16), which is expressed by means of the relative changes in both the anion and electron densities, and the $\gamma$ parameter. The condition of electrical neutrality is then performed among the density gradients, as shown in Eq. (17). Utilizing the Eqs. (16) and (17), the ratios between the gradients of the electrons density and the cations density and between the gradients of the anions density and the cations density are obtained in Eq. (18). It is noted that the ratios of these density gradients are both expressed as a function of $\gamma$ and $\alpha$. In addition, the relations between the diffusion coefficients and mobilities of different species are given in Eq. (19), based on the Einstein relation. Using the Eqs. (18) and (19), the complex ambipolar diffusion coefficient shown in Eq. (13) is firstly simplified in Eq. (20).

$$\gamma = T_e / T_i, \tag{15}$$

$$\frac{\nabla n_-}{n_-} = \gamma \frac{\nabla n_e}{n_e}, \tag{16}$$

$$\nabla n_+ = \nabla n_- + \nabla n_e, \tag{17}$$

$$\frac{\nabla n_e}{\nabla n_+} = \frac{1}{1+\gamma\alpha}, \frac{\nabla n_-}{\nabla n_+} = \frac{\gamma\alpha}{1+\gamma\alpha}, \tag{18}$$

$$\frac{D_-}{D_+} = \frac{\mu_-}{\mu_+}, \frac{D_e}{D_+} = \gamma \frac{\mu_e}{\mu_+}, \tag{19}$$

$$D_{a+} = D_+ \frac{(1+\gamma+2\gamma\alpha)\left(1+\alpha\frac{\mu_-}{\mu_e}\right)}{(1+\gamma\alpha)\left(1+\frac{\mu_+}{\mu_e}(1+\alpha)+\frac{\mu_-}{\mu_e}\right)}. \tag{20}$$

At present, the simplified ambipolar diffusion coefficient is a function of the quantities $D_+$, $\alpha$, $\gamma$, and the mobilities $\mu_+, \mu_-, \mu_e$. Then, the approximations between the mobilities of heavy cations and anions and the electron mobility, as illustrated in Eq. (21), are used to further simplify the coefficient in Eq. (22). The present article is focused on the highly electronegative plasma, and so if the electronegativity $\alpha$ is high enough, e.g., ~100, the coefficient is again simplified to be a constant, $2D_+$, as shown in Eq. (23).

$$\mu_-/\mu_e \ll 1, \mu_+/\mu_e \ll 1, \tag{21}$$

$$D_{a+} \cong D_+ \frac{1+\gamma+2\gamma\alpha}{1+\gamma\alpha}, \tag{22}$$

$$\alpha \gg 1, D_{a+} \cong 2D_+. \tag{23}$$

Still at the Boltzmann balances of the electrons and anions, the relation between the anion and electron densities is given in Eq. (24), which is different with the Eq. (16) that is focused on the density gradient. Here, $n_{e0}$ and $n_0$ are the fixed central densities of electrons and anions, respectively. In non-thermal equilibrium plasma, the electron temperature $T_e$ is high, with an amplitude of several electron volts; however, the anion temperature $T_i$ is low, which is approximately hundreds of Kelvin. Therefore, the value of $\gamma$ is of the order of 100. At such a high value of $\gamma$, the term on the right-hand side of Eq. (24), $\left(\dfrac{n_-}{n_{-0}}\right)^{1/\gamma}$, tends to one and the electrons density on the left side of it becomes spatially unchangeable, i.e., equal to $n_{e0}$. This process is expressed in Eq. (25).

$$\frac{n_e}{n_{e0}} = \left(\frac{n_-}{n_{-0}}\right)^{1/\gamma}, \tag{24}$$

$$\gamma \cong 100.0, \left(\frac{n_-}{n_{-0}}\right)^{1/\gamma} \cong 1.0, n_e \cong n_{e0}. \tag{25}$$

In Eq. (26), the cation continuity equation of the electronegative plasma before simplified is given. On the left side of the equation, the flux is expressed as the product of the ambipolar diffusion coefficient and the density gradient of the cation. On the right-hand side of the equation, the chemical sources include the ionization reactions that create the cations and the recombination reactions that deplete the cations. Here, $n_0$ is the density of the target neutral atom for the electron collisions. $K_{iz}$ and $K_{rec}$ are the rate coefficients for the ionization and recombination reactions, respectively. The simplified continuity equation is shown in Eq. (27) and the simplifications used are given in the next three aspects. Firstly, on the left side of the continuity equation, the simplified ambipolar diffusion coefficient of Eq. (23), $2D_+$, is used in the flux expression. Secondly, on the right side of the continuity equation, the simplified constant electron density of Eq. (25), $n_{e0}$, is used in the ionization chemical term. Thirdly, the recombination loss of cations is neglected in the chemical source terms, which is validated at the low pressure, as mentioned in Sec. I. The simplified cation continuity equation of Eq. (27) can be analytically solved and the solution of it is a parabolic function, as shown in Eq. (28). Herein, the $\alpha_0$ parameter defines the electronegativity of the center, as shown in Eq. (29), and $l$ is the nominal position where the electronegativity is zero. Utilizing the Eqs. (7, 25), the cations density function of Eq. (28) can be expressed into the electronegativity function, as shown in Eq. (30).

$$-\frac{d}{dx}\left(D_{a+}\frac{dn_+}{dx}\right) = K_{iz}n_0 n_e - K_{rec}n_+ n_-, \tag{26}$$

$$-2D_+ \frac{d^2 n_+}{dx^2} = K_{iz}n_0 n_{e0}, \tag{27}$$

$$\frac{n_+}{n_{e0}} = \alpha_0\left(1 - \frac{x^2}{l^2}\right) + 1, \tag{28}$$

$$\alpha_0 = n_{-0}/n_{e0}, \tag{29}$$

$$\alpha = \alpha_0 \left(1 - \frac{x^2}{l^2}\right). \tag{30}$$

### (2.3) Theory of self-coagulation

The simulations show the self-coagulation always happens to the anions at first and there are both the ambi-polar self-coagulation and the mono-polar one. The self-coagulation related theories of anions have been described in Refs. [27, 28], and herein the brief introduction of them is given. The steady-state continuity equation of anions that consists of the free diffusion flux and the negative source term given by the recombinations is expressed in Eq. (31).

$$-D_- \nabla^2 n_- = -n_- n_+ k_{rec} = -n_- v_{rec}. \tag{31}$$

Introducing the parameter, $k_- = \sqrt{\dfrac{v_{rec.}}{D_-}}$, into Eq. (31), the quasi-Helmholtz equation is obtained, as shown in Eq. (32).

$$\nabla^2 n_- - n_- \frac{v_{rec.}}{D_-} = \nabla^2 n_- - n_- k_-^2 = \nabla^2 n - n k^2 = 0. \tag{32}$$

For simplicity, the quantity subscripts are all removed in Eq. (32). In Eq. (33), this quasi-Helmholtz equation is reformed by the method of separation of variables in the cylindrical coordinate system at the azimuthal symmetry assumption.

$$\begin{aligned}
&\frac{1}{\rho}\frac{\partial}{\partial \rho}\left(\rho \frac{\partial n}{\partial \rho}\right) + \frac{\partial^2 n}{\partial z^2} - k^2 n = 0, \\
&n(\rho, z) = R(\rho) Z(z), \\
&Z'' + v^2 Z = 0, \\
&\frac{d^2 R}{d\rho^2} + \frac{1}{\rho}\frac{dR}{d\rho} - (k^2 + v^2) R = 0.
\end{aligned} \tag{33}$$

Herein, $v^2$ represents the eigenvalues. Utilizing the homogeneous boundary conditions of the above axial equation, the eigenvalues, $v^2$, and the related eigenfunctions, $Z_m(z)$, are acquired in Eq. (34).

$$\begin{aligned}
&v_m^2 = m^2 \pi^2 / l^2, \\
&Z_m = \sin(m\pi z / l), \\
&Z = \sum_{m=0}^{\infty} c_m Z_m = \sum_{m=0}^{\infty} c_m \sin(m\pi z / l).
\end{aligned} \tag{34}$$

As seen, the above radial equation is zero-order *imaginary* Bessel equation because of the negative source. Since the density value is limited at the center, the imaginary Bessel function, rather than the Hankel function, is selected. The expression for $R(r)$ is then obtained in Eq. (35).

$$R = d_m I_0(\sqrt{k^2 + v_m^2}\,\rho) = d_m I_0(\sqrt{k^2 + m^2\pi^2/l^2}\,\rho). \tag{35}$$

The expression for $n(\rho, z)$, which is a product of $R(\rho)$ and $Z_m(z)$ is given in Eq. (36).

$$\begin{aligned}n(\rho,z) = R(\rho)Z(z) &= \sum_{m=0}^{\infty} c_m \sin(m\pi z/l) \cdot d_m I_0(\sqrt{k^2 + v_m^2}\,\rho) \\ &= \sum_{m=0}^{\infty} a_m \sin(m\pi z/l) \cdot I_0(\sqrt{k^2 + m^2\pi^2/l^2}\,\rho).\end{aligned} \tag{36}$$

It is seen from Eq. (37) the delta distribution unrelated to the spatial coordinates evolves at using the mathematic limit ideas to the obtained density, which indicates the physics of self-coagulation, actually. More details about the self-coagulation theory can be found in Refs. [27, 28].

$$\begin{aligned}n(\rho,z) = R(\rho)Z(z) &= \sum_{m=0}^{\infty} a_m \sin(m\pi z/l) \cdot I_0(\sqrt{k^2 + m^2\pi^2/l^2}\,\rho) \\ &= \lim_{m\to\infty}\left[a_m \sin(m\pi z/l)\cdot\infty\right] = \lim_{m\to\infty}\left[a_m \sin(m\pi z/l)\cdot\lim_{z\to 0}\frac{1}{z}\right] \\ &= \lim_{z\to 0}\left[\lim_{m\to\infty} a_m \sin(m\pi z/l)\cdot\frac{1}{z}\right] = \lim_{z\to 0}\left[\lim_{m\to\infty} a_m \cdot\frac{\sin(m\pi z/l)}{z\pi/l}\cdot\frac{\pi}{l}\right] \\ &= \lim_{\zeta\to 0}\left[\lim_{m\to\infty} a'_m \cdot\frac{1}{\pi}\frac{\sin(m\zeta)}{\zeta}\right] = \lim_{\zeta\to 0}\left[a'_\infty \lim_{m\to\infty}\frac{1}{\pi}\frac{\sin(m\zeta)}{\zeta}\right] \\ &= a'_\infty \lim_{\zeta\to 0}\delta(\zeta).\end{aligned} \tag{37}$$

### III. Results and analysis
### (3.1) Stratification, parabola core, and ambi-polar self-coagulation

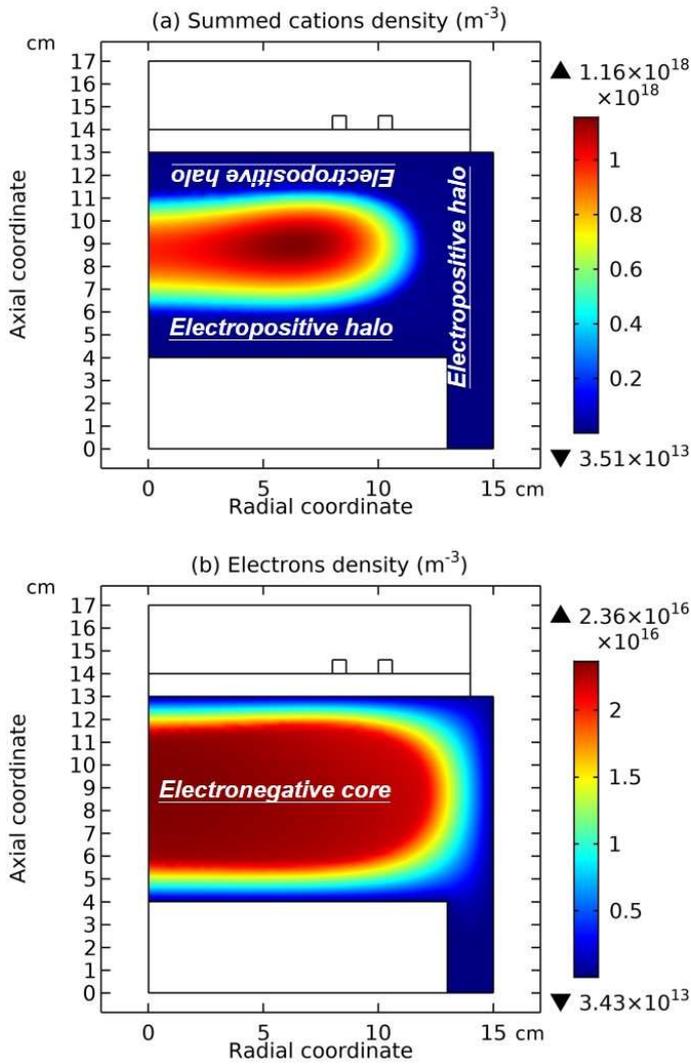

Figure 1. Simulated two-dimensional profiles of the summed cations density (a) and the electrons density (b), by the fluid model of Sec. (2.1). The discharge conditions of Ar/$SF_6$ ICP are 300W, 10mTorr and 10% $SF_6$ content and the simulated time is $0.1\,\text{s}$.

The simulated two-dimensional profiles of the summed cations density and the electrons density by the fluid model of Sec. (2.1) are presented in Fig. 1, and the axial and radial density profiles of the cations, the anions (also summed), and the electrons are given in Fig. 2. The discharge conditions of the Ar/$SF_6$ ICP source are 300 W, 10mTorr and 10% $SF_6$ content, and the simulated time is 0.1 s. It can be seen in Fig. 1 that the entire discharge structure is stratified into an electronegative core and an electropositive halo, and in the electronegative core the density axial profiles of the summed cations and anions in Fig. 2(a) are both parabolic, as predicted by the analytical theory of Sec. (2.2). For better demonstrating such a parabola characteristic, the density axial profiles of cations and anions simulated are still compared to a parabola function constructed and based on the two critical data points that are sampled from the simulated density curve of summed cations, in Fig. 3. As seen, the similarity between the simulated and predicted parabolas is satisfactory. The simulated electrons density of Fig. 1(b) decreases in the electropositive halo, and is constant in the electronegative core, again in accord to the theoretical prediction of Sec. (2.2)

in Eq. (25). The simulated and constant electron density in the electronegative core validates the Boltzmann balance approximation used for the electrons and anions therein in the theory of Sec. (2.2). In addition, the cations and anions densities are not monotonic in their radial profiles but the density bumps appear in the core, as shown in Figs. 1(a) and 2(b). In Fig. 4(a, b) the simulated net source of anions is observed to be negative in the location where the anions density is peaked, and in Fig. 4(c) the simulated plasma potential is flat in the core, implying the free diffusion of anions therein. Since the required conditions, negative chemical source and free diffusion, are satisfied, the anions are self-coagulated to form the density bump in the radial profile of them, according to the theoretical analysis of Sec. (2.3). The density bump in the radial profile of cations is given by the ambi-polar coagulation of electronegative plasma at the high electronegativity condition, like the ambi-polar diffusion of electropositive plasma that keeps the electrical neutrality. The simulated flat potential of core in Fig. 4(b) is given by the flat electrons density simulated therein in Fig. 1(b), as determined by the Boltzmann relation of electrons.

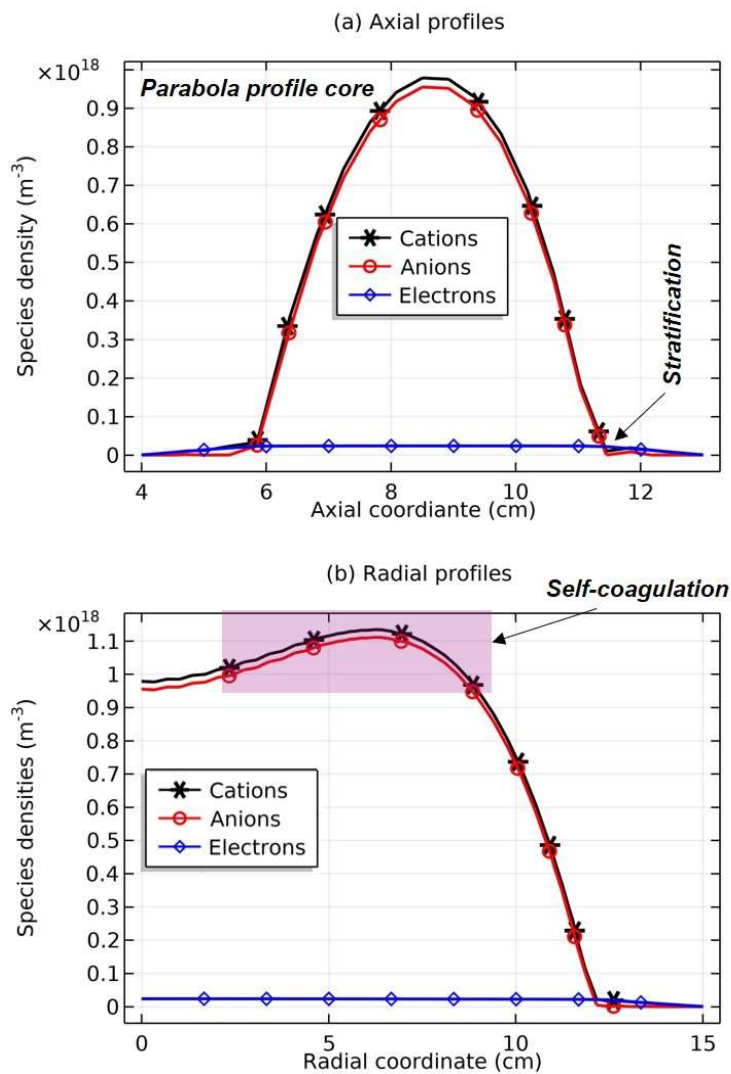

Figure 2. Simulated axial (a) and radial (b) profiles of the cations, anions, and electrons densities by the fluid model of Sec. (2.1). The discharge conditions of Ar/$SF_6$ ICP are 300W, 10mTorr and 10% $SF_6$ content and the simulated time is $0.1\,\text{s}$. In Panel (a) the density axial profiles of three plasma species are sampled along the central discharge axis, and in Panel (b) the density radial profiles of three plasma species are sampled at an axial position of $z = 8.5$ cm.

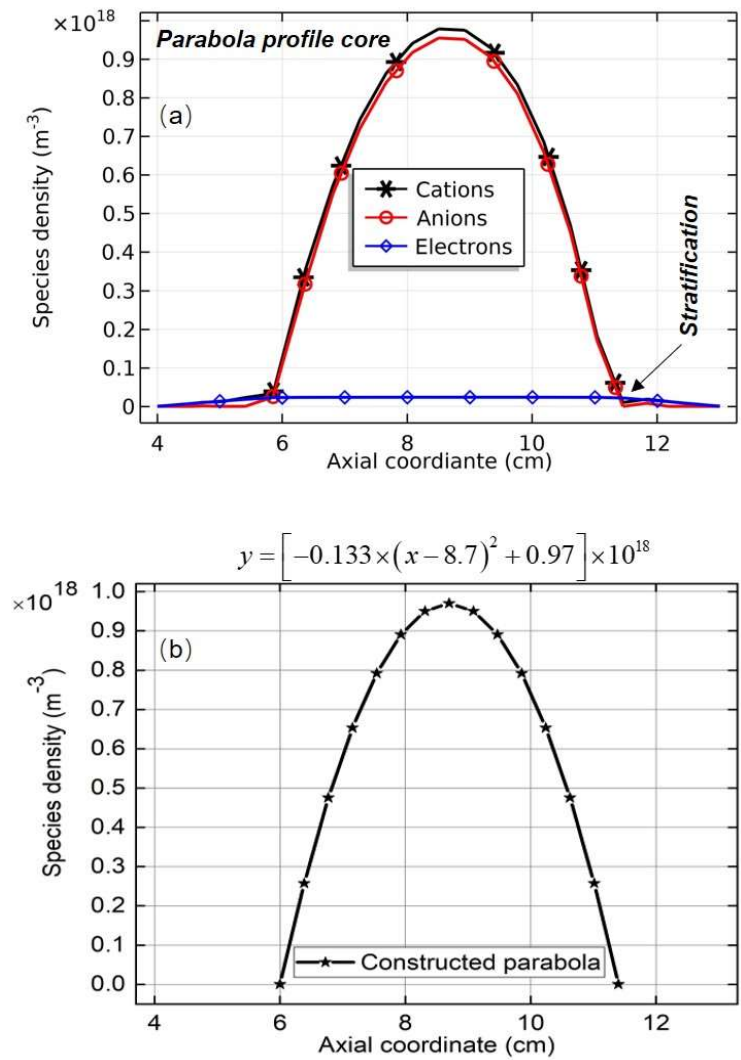

Figure 3. (a) Simulated axial profiles of species density and (b) constructed parabola function based on two critical points sampled from the simulated cations density curve, i.e., the peaked point with its coordinates, $(8.7, 0.97 \times 10^{18})$, and the truncated close-zero point with its coordinates, $(11.4, 0)$.

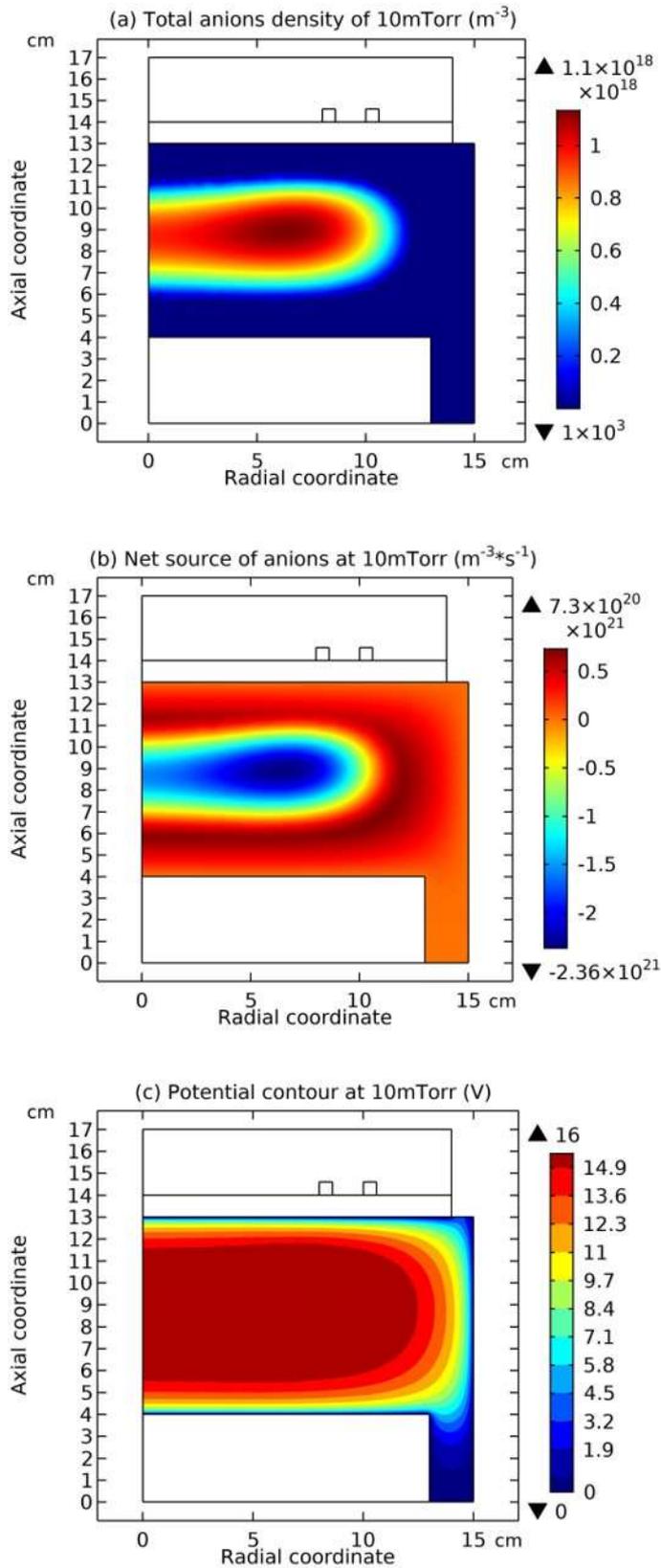

Figure 4. Simulated two-dimensional profiles of total anions density (a), net source of anions (b), and the plasma potential (c) by the fluid model of Sec. (2.1) in the Ar/SF$_6$ ICP, at 10 mTorr. The other discharge conditions and the simulated time are the same as in the Figs. 1 and 2.

## (3.2) The Boltzmann balance of anions and its possible origin, self-coagulation

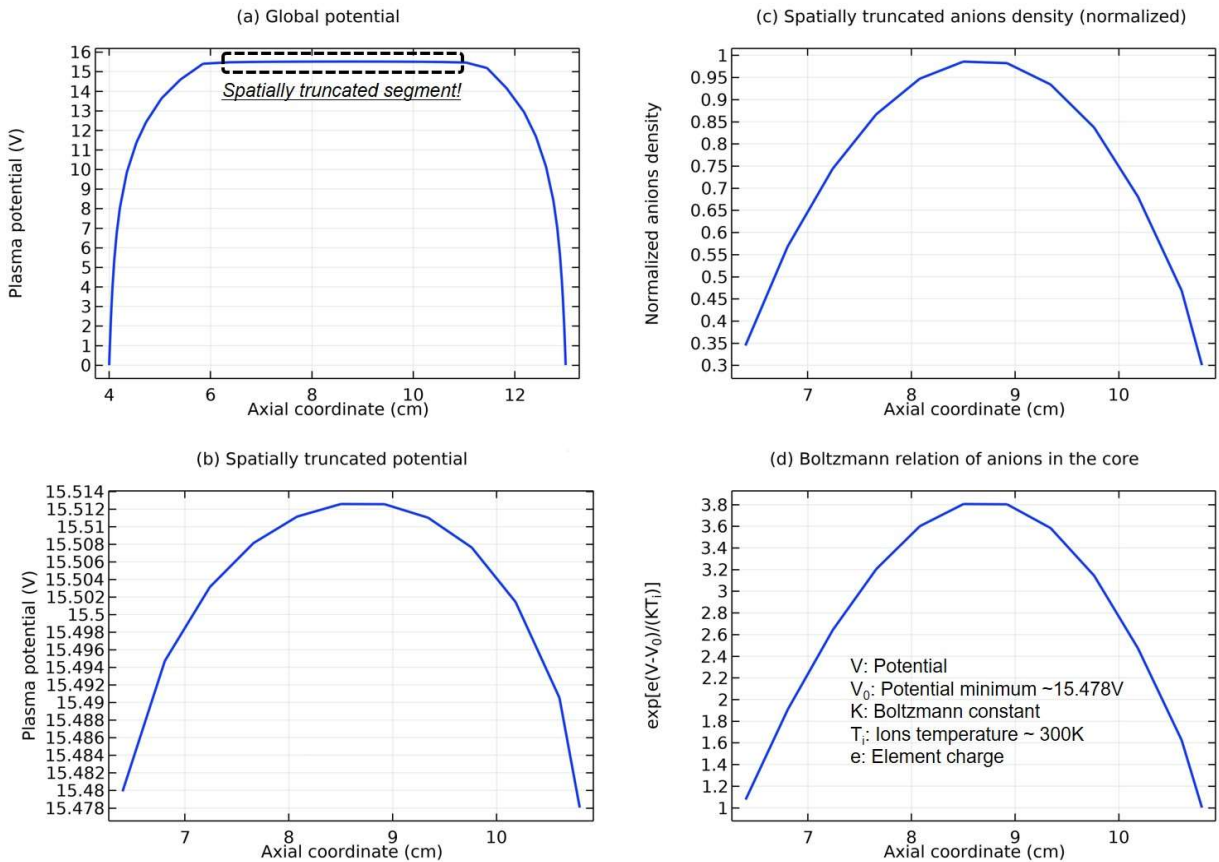

Figure 5. Simulated global and axial electrical potential of plasma (a), zoom in exhibition of the path of spatially truncated plasma potential (b), spatially truncated and normalized anions density of the same path (c) and the Boltzmann relation of anions in the path (d). The discharge conditions of the Ar/SF$_6$ ICP and the simulated time are the same as in the Figs. 1 and 2.

As mentioned before, since the electrons density is in the Boltzmann balance, the electrical potential of the plasma that is plotted globally and axially in Fig. 5(a) has the similar distribution as the electrons density shown in Fig. 1(b), that is, constant in the core but continually decreasing in the halo in a large potential unit, Volt, as observed from the vertical axis. As zooming in the potential and seen from the Figs. 5(a,b), in a truncated path as selected, the originally constant electrical potential in the core now has its definite profile, as demonstrated in a small range of potential, i.e., from $15.478\,V$ to $15.514V$. In Fig. 5(c), the normalized anions density in this path is plotted, i.e., $\dfrac{n_-}{n_{-0}}$, and in Fig. 5(d) a certain function, $\exp\left[e(V-V_0)/KT_i\right]$, is constructed based on the tiny change of the potential as shown in Fig. 5(b) and on the room temperature of anions as assumed, that is, $T_i = 300\,K$. Here, $V_0$ is the potential minimum, $K$ is the Boltzmann constant, and $e$ is the unit element charge, as illustrated in Fig. 5(d). As compared, the similarity between the two profiles of Fig. 5(c, d) indicates the anions satisfy the Boltzmann balance, i.e.,

$$\frac{n_-}{n_{-0}} \sim \exp\left[e(V-V_0)/KT_i\right].$$

From the electromagnetics, it is known the potential and energy are exchangeable through the element charge. Therefore, the tiny potential change of the path shown in Fig. 5(b) can be described by the thermodynamic unit, Kelvin. The concrete processes are $\Delta = 15.514 - 15.478 = 0.036$ V and $0.036 \text{ eV} \sim \frac{0.036 \times 1.6 \times 10^{-19}}{1.38 \times 10^{-23}} \sim 417$ K. In the formulae, the symbol, eV, is called the electronvolt. It is an energy unit, responsible for turning the potential into the energy. The electron charge, $1e = 1.6 \times 10^{-19}$ C, and the Boltzmann constant, $K: 1.38 \times 10^{-23}$ J/K, are utilized in the above formula to gives rise to the total variation of potential energy in that selected path of the core, about $417$ K. It is seen that the potential in the Boltzmann balance of anions is in the room-temperature range, which is hence a weak potential barrier.

It is noted that the Boltzmann balance of anions given by the fluid model simulation here was already validated by the experiments in Ref. [16] and widely used in the early analytical works of Refs. [11, 17-20, 22]. However, the origin of it is not clear yet. Herein, by means of the dimensional analysis of drift flux convergence and self-coagulation formulae, and by means of the characteristic of derivative of delta type density profile representing the coagulated structure, it is believed that the self-coagulation plays the role of drift that establishes the Boltzmann balance of species, at the help of free diffusion.

$$\vec{\Gamma}_d = -\mu_- n_- \vec{E}_{eff}. \tag{38}$$

$$\nabla \cdot \vec{\Gamma}_d = -\mu_- \nabla n_- \cdot \vec{E}_{eff} - \mu_- n_- \nabla \cdot \vec{E}_{eff} = -\mu_- \nabla n_- \cdot \vec{E}_{eff} - \mu_- n_- \frac{\rho_{eff}}{\varepsilon_0}. \tag{39}$$

$$\nabla n_- \cdot \vec{E}_{eff} = \nabla \delta(\vec{r}) \cdot \vec{E}_{eff} = \begin{cases} 0, & \text{for } r \neq 0 \\ non-existing, & \text{for } r = 0 \end{cases}. \tag{40}$$

$$[\mu_-] = \frac{m^2}{V \cdot s}, \ [n_-] = \frac{1}{m^3}, \ [\rho_{eff}] = \frac{C}{m^3}, \ [\varepsilon_0] = \frac{C}{m \cdot V},$$

$$[\mu_- n_- \frac{\rho_{eff}}{\varepsilon_0}] = \frac{1}{m^3 \cdot s} = [-n_- \nu_{rec}]. \tag{41}$$

$$\begin{aligned} &-D_- \nabla^2 n_- - (-n_- \nu_{rec}) = 0, \\ &-D_- \nabla^2 n_- - \mu_- n_- \nabla \cdot \vec{E}_{eff} = 0. \end{aligned} \tag{42}$$

In Eq. (38), the drift flux of anions is assumed based on an *effective* electric field, and in Eq. (39) the divergence of such a flux is calculated, which is divided into two parts. The first part correlates the derivative of anions density, which will either be zero or not exist, at the assumption of delta profile, as shown in Eq. (40). For the second part of flux divergence, the Poisson equation is used to transfer it to be a product of anions density, $n_-$, and an *effective* charge density, $\rho_{eff}$, together with the two constants, $\mu_-$, $\varepsilon_0$, as shown at the end of Eq. (39). After this transform, the expression of the second part of drift flux divergence is like that of the negative source in Eq. (31) of Sec. (2.3). The dimensional analysis of Eq. (41) further verifies the similarity between the second part of drift flux divergence and the negative source. So, the Quasi-Helmholtz equation in Eq. (31) is rewritten here in Eq. (42), and the negative source of it correlates directly to an *effective* drift flux. It is stressed that only the medium of flux is self-coagulated and the first part of drift flux divergence attenuates, this correlation exists, which implies that the self-coagulation provides the precursor of driving force for the electronegative plasma to evolve out the Boltzmann balance of anions, i.e., the real and weak potential barrel shown in Fig. 5(b).

## (3.3) Investigation on the double layer
## (a) Its transferred models

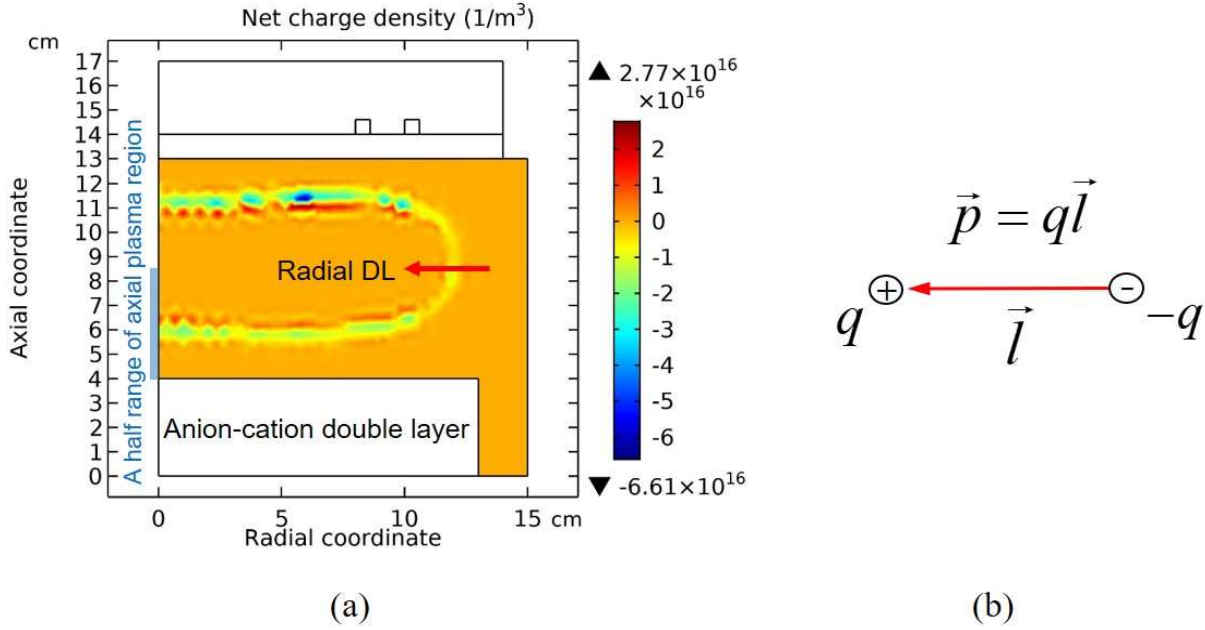

Figure 6. (a) Simulated two-dimensional profile of net charge density, by the fluid model of Sec. (2.1) in the Ar/SF$_6$ ICP, and (b) transferred dipole model of double layer, abbreviated as DL, in the radial direction. The discharge conditions and the simulated time are the same as in the Figs. 1 and 2. The blue bar and corresponding text, a half region of axial plasma region, in Panel (a) will be used in Fig. 10(a).

  Besides for the stratification, the core parabola profile, and the coagulation, the self-consistent fluid simulation still shows that a double layer appears at the interface of stratification, which agrees with the simple fluid model prediction that includes the Poisson equation in Refs. [17-20, 22]. In Fig. 6(a), the simulated two-dimensional profile of net charge density in the Ar/SF$_6$ ICP is shown. Upon comparing Fig. 6(a) to Fig. 1, it is seen that at the interface of the electronegative core plasma and the electropositive halo plasma, the double layer structure, i.e., one layer of negative charge adjacent to one layer of positive charge, is appeared. In Fig. 6(b), this double layer is modelled as a dipole along or anti- along the direction of plasma transport, i.e., from the core/halo to halo/core. Besides, it is seen from the figure that the two layers of charge are very close to each other. So, in Fig. 7, the spatial distribution of the electric field intensity of such a dipole at the limit of dipole distance tending to zero is given. It is seen that the field is strong at the dipole center but tends to zero at other points. The concrete mathematic process is presented in Eqs. (43-46). In both the Fig. 7 and the related equation set, the point, $O$, is the dipole center and the point, $A$, is an arbitrary *far-field* point of the space. $E_O$, $E_A$ are the dipole fields of points, $O$, $A$, respectively. Herein, $l$ is the distance of dipole moment and $q$ is the dipole charge. Hence, $\vec{p} = q\vec{l}$ and it is the dipole moment. $r$ is the distance of dipole center to the point, $A$, and $\vec{r}_0$ is the corresponding unit vector. $\varepsilon_0$ is the vacuum permittivity. The field intensities at *far-field* points tend to zero because of the counteracting effect of positive and negative charges while the field intensity at the dipole center is infinitely large because of the localization effect, at the limit of $l \to 0$.

$$E_O = \frac{2q}{\pi\varepsilon_0 l^2}, \tag{43}$$

$$\lim_{l \to 0} E_O = \lim_{l \to 0} \frac{2q}{\pi\varepsilon_0 l^2} = \infty, \tag{44}$$

$$\vec{E}_A = \frac{1}{4\pi\varepsilon_0 r^3}\left[-\vec{p} + 3(\vec{r}_0 \cdot \vec{p})\vec{r}_0\right], \tag{45}$$

$$l \to 0, \vec{p} = q\vec{l} \to \vec{0}, \Rightarrow \vec{E}_A \sim \vec{0}. \tag{46}$$

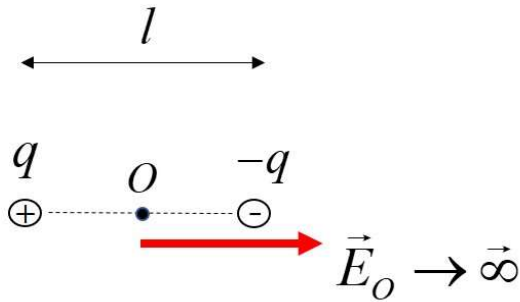

(a) Electric field intensity of dipole moment at the dipole center

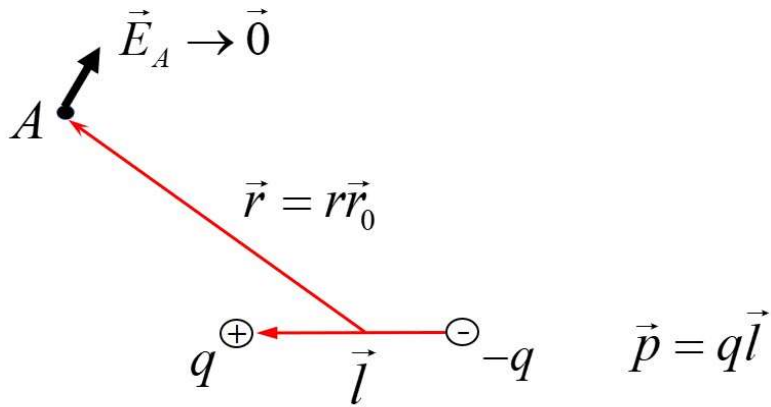

(b) Electric field intensity of dipole moment at arbitrary *far-field* point of the space

Figure 7. Spatial distribution of the electric field intensity of the transferred dipole model at the limit of $l \to 0$ at (a) the dipole center and (b) arbitrary *far-field* point. Herein, $l$ is the distance of dipole moment.

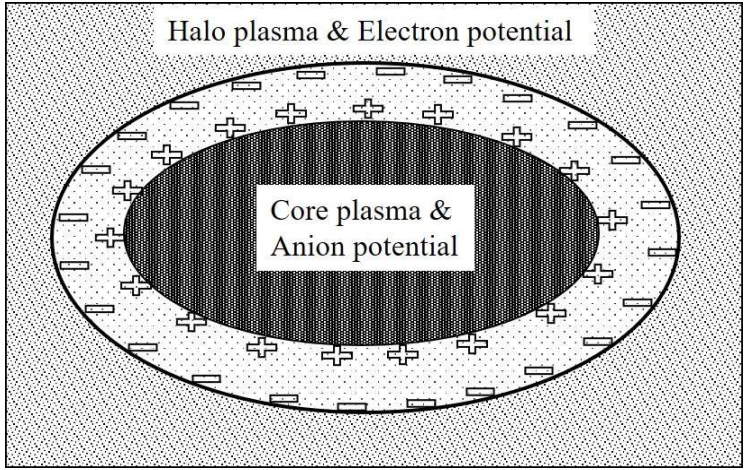

Figure 8. Transferred ellipsoid capacitor model of double layer

Considering the three-dimensional geometry of cylindrical chamber and the axial symmetry, the double layer, when seen along the whole interface of the electronegative core plasma and the electropositive halo plasma, can be transferred to a capacitor of the ellipsoid, as shown in Fig. 8. Herein, the electron potential in the external halo plasma is meant the ambi-polar diffusion potential in a unit of electronvolt, and the anion potential in the internal core plasma is meant the weak potential in a unit of Kelvin that supports the Boltzmann balance of anions in Fig. 5 of Sec. (3.2). Different with the CCP source that is radio frequency vibrated, the transport of ICP source is more like a direct current (DC) plasma source in the radial and axial directions. The capacitor is known to be able to block the direct current, which means that the electropositive halo is separated from the electronegative core, as seen from Fig. 9 where a transferred circuit model of such an ICP source is given. The stratification of the Ar/$SF_6$ ICP discharge structure through the capacitor model of the double layer is triggered by the external power source, as illustrated in this figure. As seen next, the electrical field and potential properties of the transferred dipole and capacitor models of double layer are both validated by the simulation.

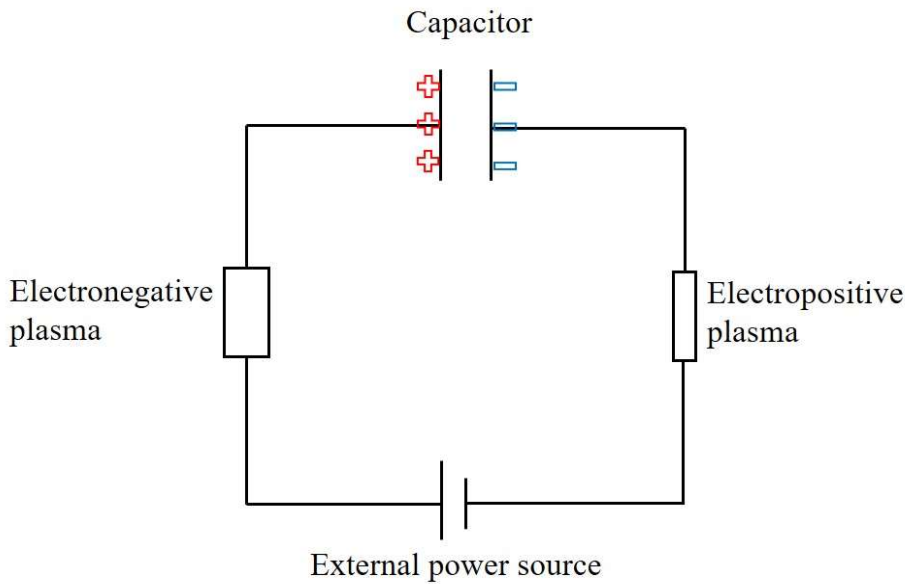

Figure 9. Transferred circuit model of the Ar/$SF_6$ ICP.

## (b) Double-valued properties of plasma edge potential and cations flux, at the DL

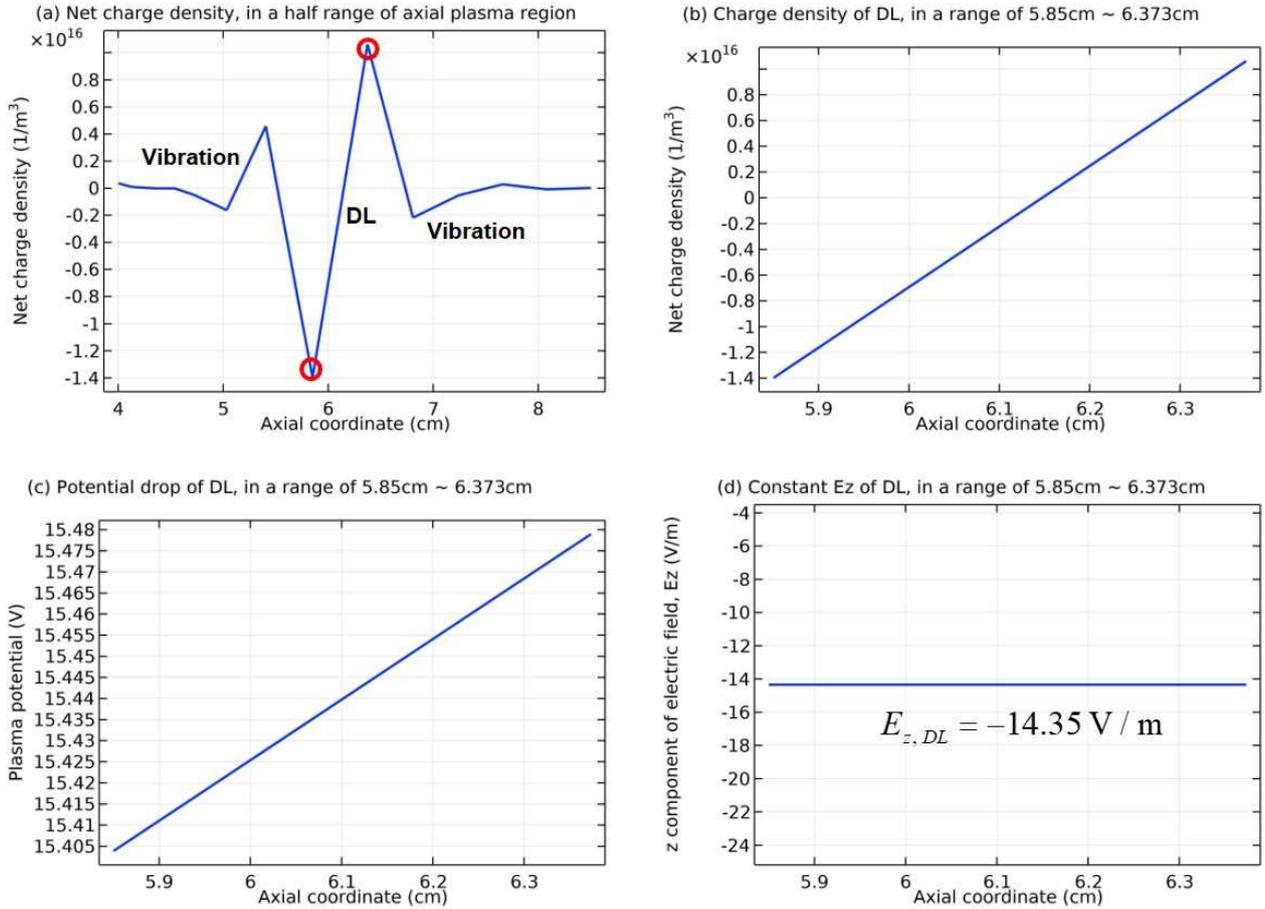

Figure 10. (a) Net charge density in a half range of axial plasma region (see Figure 6(a) for reference), (b) charge density of double layer in an axial range from 5.85 to 6.373 cm, (c) potential drop of double layer still in this range, and (d) display for the constant axial electrical field component of double layer, $-14.35 \text{ V}/\text{m}$, in the same range, all along the central axis of discharge and given by the fluid simulation of Sec. (2.1). The discharge conditions of the Ar/SF$_6$ ICP and the simulated time are the same as in the Figs. 1 and 2. In Panel (a), the region of DL is marked and at the two sides of it, the ionic and acoustic vibrations appear, which will be analyzed in Sec. (3.3c)

In this section, the terms, plasma edge potential and flux, were illustrated in the literature review of Sec. I. They were defined in the early analytical works of Sheridan *et al* in Refs. [19, 20, 22]. In these works, both the method based on the plasma approximation, i.e., without the Poisson equation, and the one that includes the non-neutral region, i.e., with the Poisson equation, were used. The applications of the two methods and the correlation of them revealed that at the plasma edge, i.e., the interface between the electrically neutral plasma and the non-neutral region, both the potential and the flux of cations are double-valued, and when the flux magnitude of cations with the high edge potential was larger than the one with the low edge potential, the double layer appeared at the plasma edge. Our fluid simulation perfectly describes this picture in Figs. 10-12, and moreover, it finds out the behind physics for explaining such a trend related to the double layer in the electronegative plasma sources.

In Fig. 10(a), the simulated net charge density by the fluid model of Sec. (2.1), in a half range of axial plasma region along the central axis of discharge, is given, where the double layer and the related vibrations are presented. In Fig. 10(b,c,d), the net charge density, the potential drop, and the axial electric field component in an axial range of double layer location, i.e., from 5.85 to 6.373 cm, are presented. As seen before, the double layer in Fig. 10(b) consists of two charge layers with different polarities. The potential in Fig. 10(c) drops with the direction oriented to the halo and it, called as the plasma edge potential, is indeed double-valued in this non-neutral region, in accord to the theoretical predictions of Refs. [19, 20, 22]. The axial electrical field of double layer in Fig. 10(d) is constant and its magnitude is $-14.35$ V/m , which agree qualitatively with the dipole and capacitor model predictions on the electric field properties in Sec. (3.3a). In Fig. 11, the general trends of the axial electrical field and the potential for the regions of the sheath, the electropositive halo, the double layer, and the electronegative core, are presented. From this figure, the intensities of electrical field in the different regions can be ordered, i.e., sheath, electropositive halo, double layer, and electronegative core. Similarly, the potential is ordered as well, but reversely as noticed, i.e., electronegative core, double layer, electropositive halo, and sheath. In a word, the potential consecutively decreases from the inner core to the outer sheath.

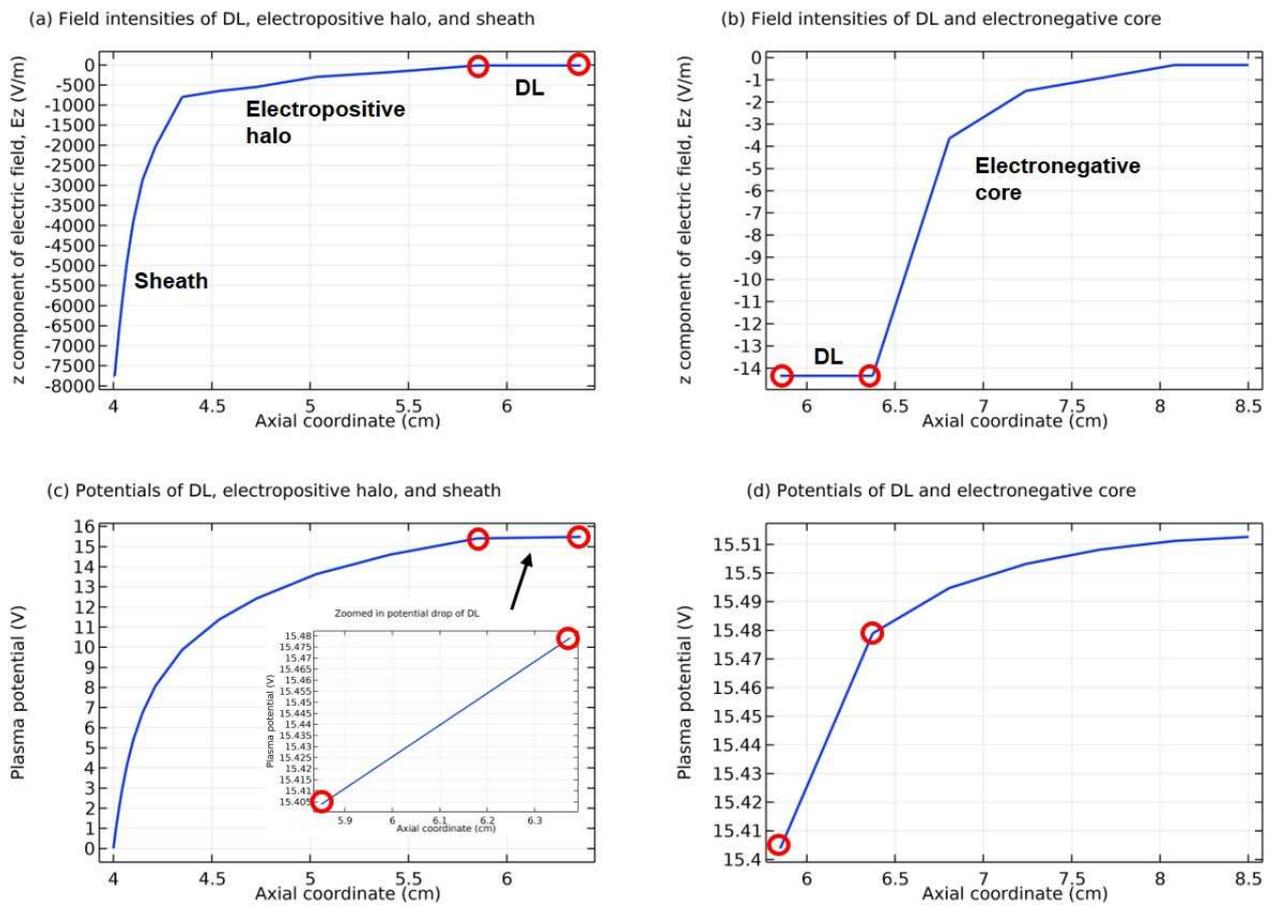

Figure 11. Axial profiles of axial electrical field component for (a) the regions of double layer, electropositive halo, and sheath and for (b) the regions of double layer and electronegative core, and still the axial profiles of potential for (c) the regions of double layer, electropositive halo, and sheath and for (d) the regions of double layer and electronegative core, all along the central axis of discharge and given by the fluid simulation of Sec. (2.1). The discharge conditions of the Ar/SF$_6$ ICP and the simulated time are the same as in the Figs. 1 and 2. In Panels (a) and (b) the electrical field intensity of double layer is truly constant as demonstrated in Fig. 10(d), and in Panel (c) the potential of double layer is not flat as illustrated in the inserted sub-figure of Panel (c) and in Panel (d) when the small ranges of potential value are used for the vertical axes.

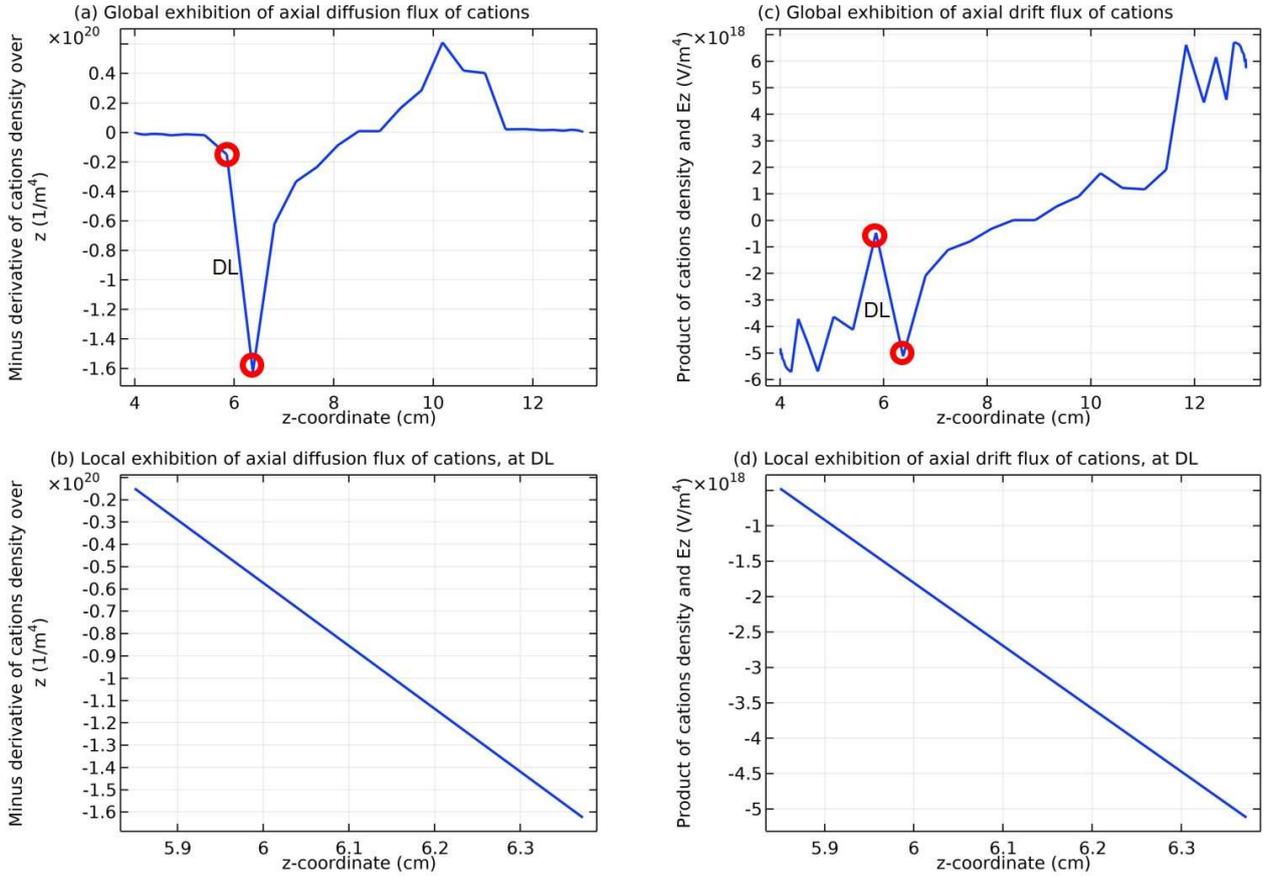

Figure 12. (a) Global exhibition of axial *diffusion* flux of cations and (b) local exhibition of it at the double layer, and (c) global exhibition of axial *drift* flux of cations and (d) local exhibition of it at the double layer, all along the central axis of discharge and given by the fluid simulation of Sec. (2.1). The discharge conditions of the Ar/SF$_6$ ICP and the simulated time are the same as in the Figs. 1 and 2. The definitions for the diffusion and drift fluxes herein are as follows, i.e., the axial *drift* flux of cations is represented by the product of cations density and axial electrical field and the axial *diffusion* flux is represented by the minus derivative of cations density over the axial coordinate. See the dimensions of the two representative fluxes for reference. The qualitative trends of two fluxes, not the quantitative values of them, are emphasized in this figure.

The axial profiles of representative axial diffusion and drift fluxes of cations are plotted both globally in the whole range in Fig. 12(a, c) and locally in the small range of double layer in Fig. 12(b, d). In Fig. 12(a, c) the positions of double layers in the flux curves are marked and both the diffusion and drift fluxes of cations are double-valued in the range of double layer, again in accord to the theoretical predictions of Refs. [19, 20, 22]. More clearly in Fig. 12(b, d) the magnitudes of the two types of cations fluxes near the electronegative core, with the high edge potential as seen from Figs. 10(c) and 11(d), are indeed larger than the ones near the electropositive halo, with the low edge potential as seen from Figs. 10(c) and 11(c), again in agreement with the theoretical predictions of Refs. [19, 20, 22] on the relation of plasma edge potential and flux that are both double-valued. It is originated from the simulated facts that the potential consecutively drops from the core to the halo, through the double layer, as shown in Fig. 11(c,d) and the flux magnitudes in the core are naturally larger than the ones in the halo where the density of cations summed is several orders lower, as shown in Fig. 2(a).

**(c) Analysis on the double layer origin**

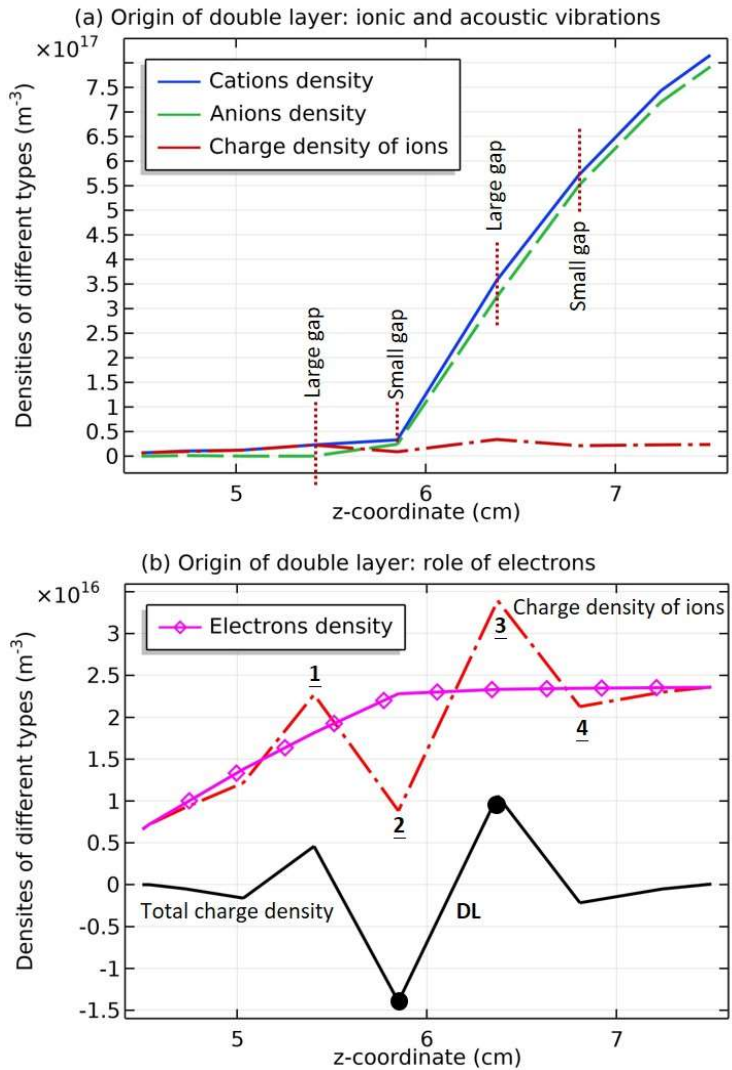

Figure 13. (a) Axial profiles of cations density (blue line), anions density (green line) and charge density of ions (red line), i.e., the deduct of cations and anions densities, and (b) axial profiles of charge density of ions (red line), electrons density (purple line) and the total charge density (black line), i.e., the deduct of cations density and the sum of anions and electrons densities, all along the central axis of discharge and given by the fluid simulation of Sec. (2.1). The discharge conditions of the Ar/$SF_6$ ICP and the simulated time are the same as in the Figs. 1 and 2.

In Fig. 13, the axial profiles of cations density, anions density, charge density of ions, electrons density and the total charge density simulated are shown. It is seen in Fig. 13(a) that the double layer hidden in the curve of charge density of ions is originated from the ionic and acoustic vibrations of cations and anions densities, and in Fig. 13(b) that the electrons are less related to the formation of double layer and the role of them is to just shift down the curve of ionic charge density. The double layer marked in Fig. 13(b) is formed at the interface between the core and halo, where the ionic and acoustic vibration is the strongest. Away from the interface, these vibrations are smoothed by the diffusion of ionic species, as illustrated in Ref. [42]. It is concluded that although the vibrations, or oscillations, are smoothed by diffusion both in the core and halo, the double layer is formed by the oscillations of ionic species at the interface, where the diffusion is temporally blocked by the border effect of two different plasma types therein. Hence, it belongs to the acoustic double layer type as defined in Ref. [31-33].

### (3.4) Temporal evolution of discharge in the simulation
### (a) Relative independence between self-coagulation and double layer

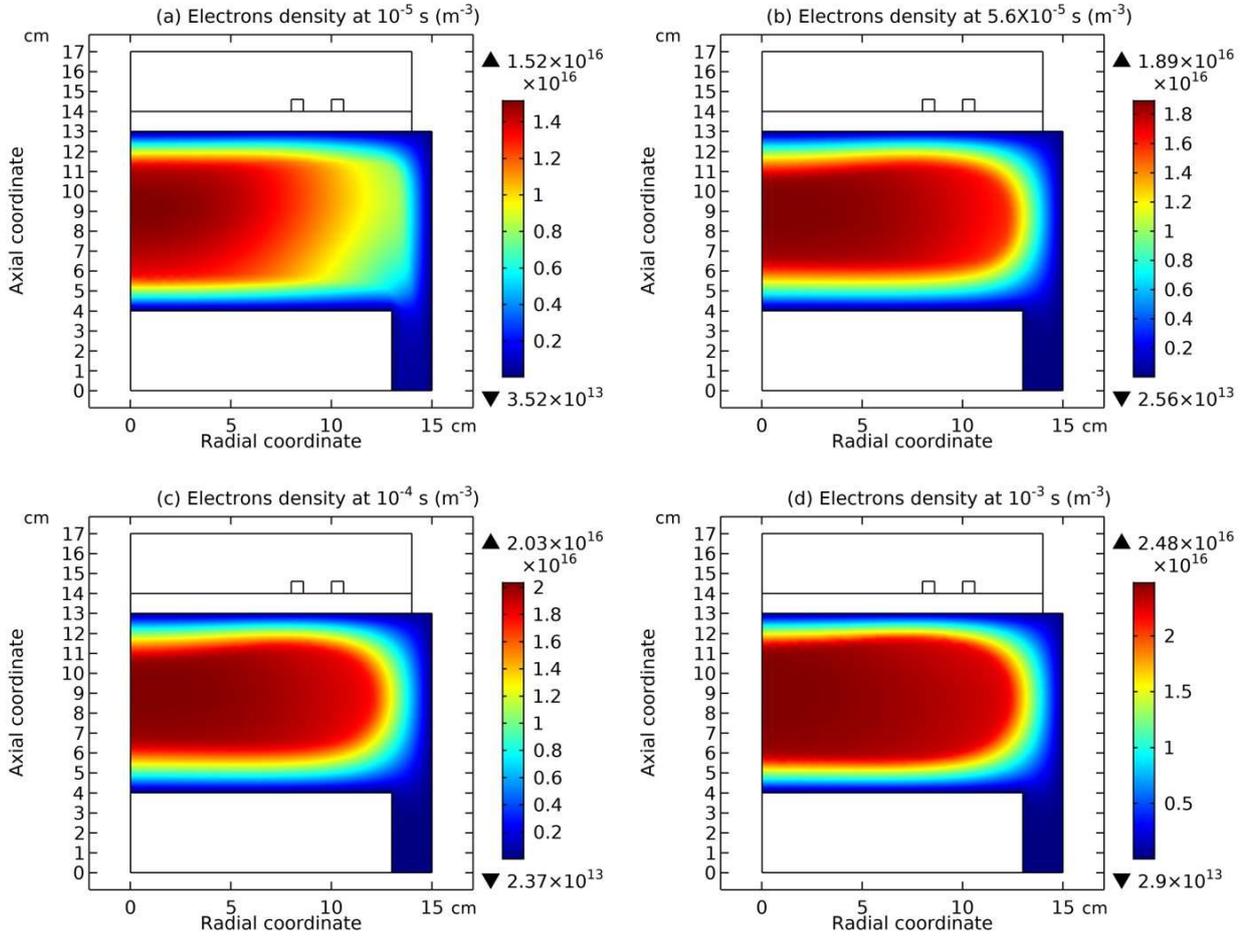

Figure 14. Simulated two-dimensional profile of electrons density at (a) $10^{-5}$ s, (b) $5.6 \times 10^{-5}$ s, (c) $10^{-4}$ s, and (d) $10^{-3}$ s, by the fluid model of Sec. (2.1). The discharge conditions of the Ar/SF$_6$ ICP are the same as in Figs. 1 and 2.

In Figs. 14-17, the temporal evolutions of two-dimensional profiles of electrons density, total anions density, net source of anions, and net charge density in the fluid simulation of Sec. (2.1) are given. It is seen that the discharge experiences two stages in total, i.e., the push by means of external ambi-polar potential barrel, e.g., Figs. 14(a), 15(a), 16(a), and 17(a) where the borders between the core and halo are stiff, and the internal evolution of core by means of self-coagulation, e.g., Figs. 14(d), 15(d), 16(d), and 17(d) where the borders between the core and halo become soft. The evolution and final formation of anions negative source trigger the self-coagulation, as illustrated in Figs. 16(b,c,d). It is seen from Figs. 14-16 that the core has its own evolution, i.e., by means of the self-coagulation, the anions Boltzmann balance, the parabola profiles of ionic densities, and the uniform electrons density profile are gradually achieved. The double layer at the interface between the core and halo is relatively independent on the self-coagulation. It appears before the self-coagulation is formed, as seen from Fig. 17(a). The double layer, representing the border of core and halo, is softened when the core starts its internal gentle evolution. The relative independence between the double layer and self-coagulation validates the analysis on the double layer origin in Sec. (3.3c), i.e., ionic and acoustic vibrations based on the pre-push of ambi-polar potential that accumulates net anions at first. It is stressed that at the low pressure, i.e., 10mTorr, the double layer is formed by the ionic and acoustic vibrations of net anions density accumulated. When the pressure increases, e.g., to 90mTorr, it is seen that the blue sheath and instability, not the double layer, are formed, as illustrated in the Figs. 22-24 of Sec. (3.5).

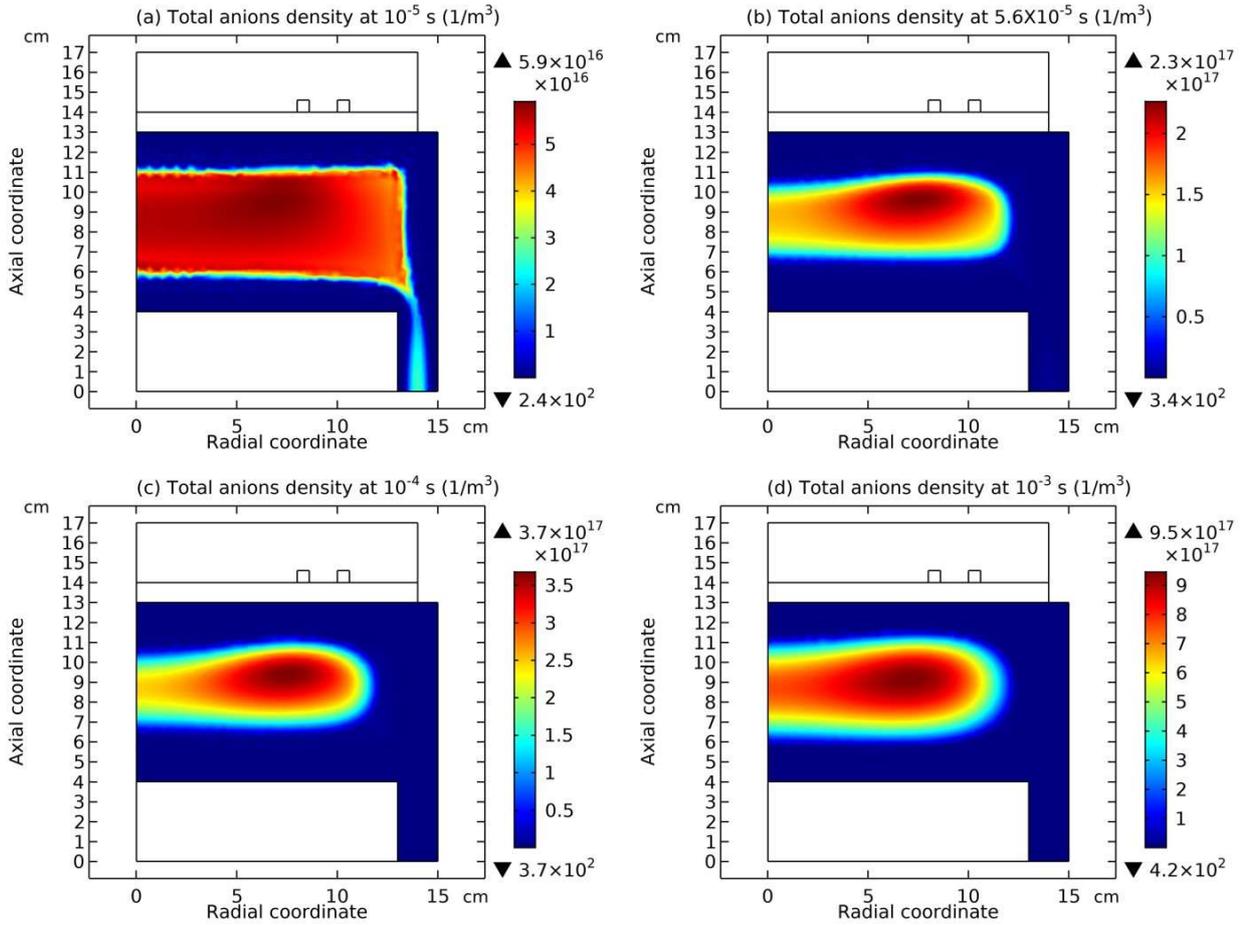

Figure 15. Simulated two-dimensional profile of total anions density at (a) $10^{-5}$ s, (b) $5.6 \times 10^{-5}$ s, (c) $10^{-4}$ s, and (d) $10^{-3}$ s, by the fluid model of Sec. (2.1). The discharge conditions of the Ar/SF$_6$ ICP are the same as in Figs. 1 and 2.

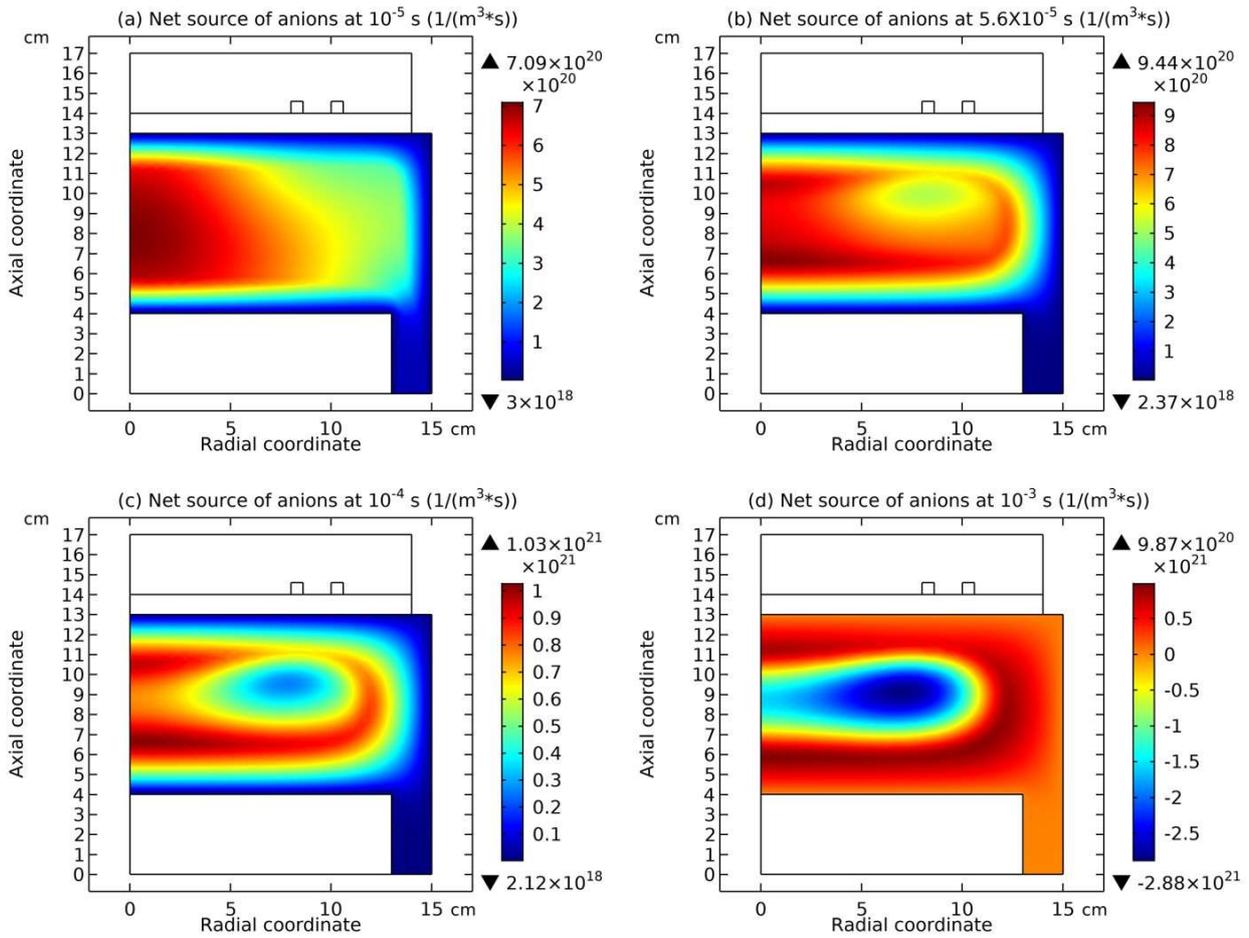

Figure 16. Simulated two-dimensional profile of net chemical source of anions at (a) $10^{-5}$ s, (b) $5.6 \times 10^{-5}$ s, (c) $10^{-4}$ s, and (d) $10^{-3}$ s, by the fluid model of Sec. (2.1). The discharge conditions of the Ar/SF$_6$ ICP are the same as in Figs. 1 and 2.

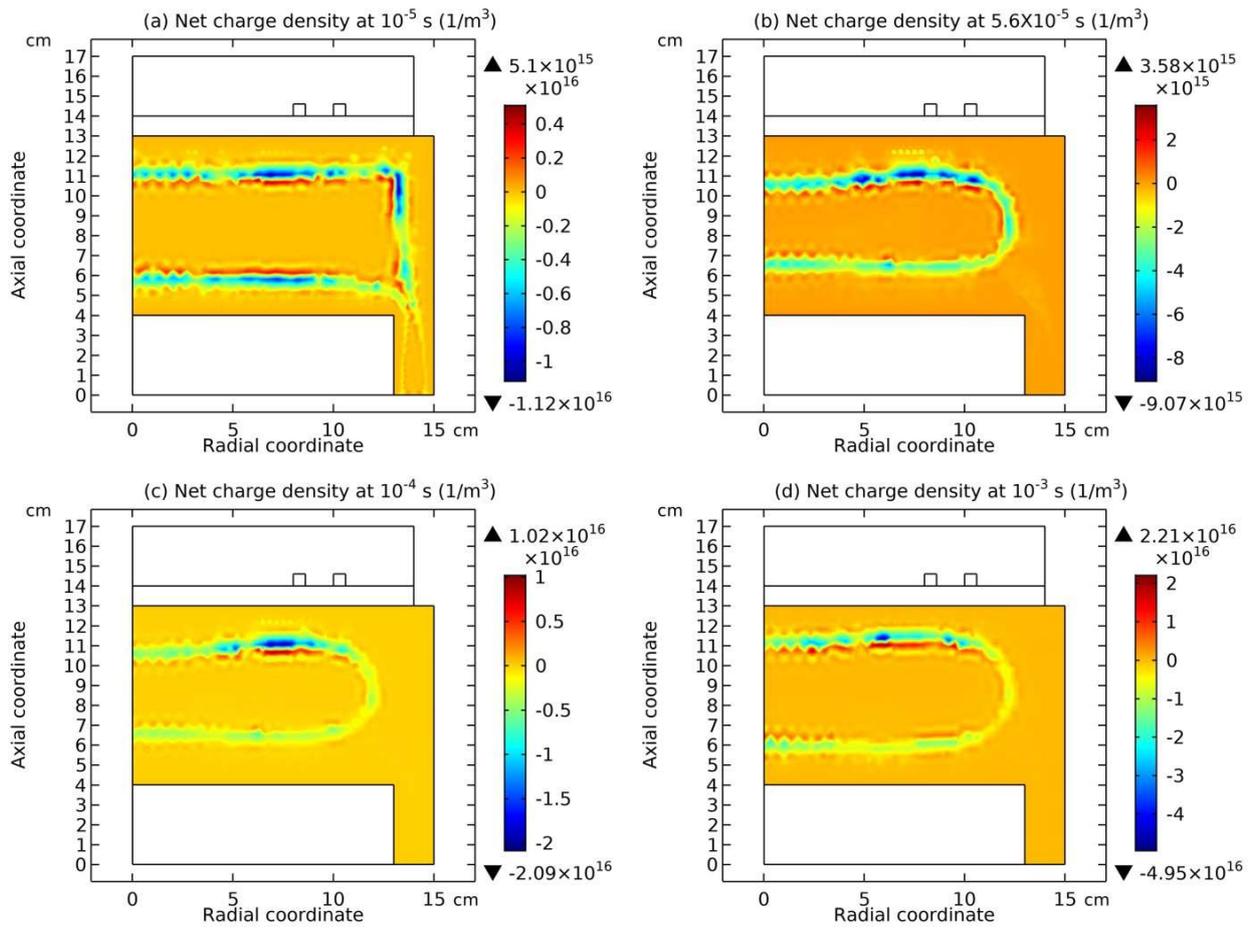

Figure 17. Simulated two-dimensional profile of net charge density at (a) $10^{-5}$ s, (b) $5.6 \times 10^{-5}$ s, (c) $10^{-4}$ s, and (d) $10^{-3}$ s, by the fluid model of Sec. (2.1). The discharge conditions of the Ar/SF$_6$ ICP are the same as in Figs. 1 and 2.

## (b) Convergence of simulation and harmony among multiple processes

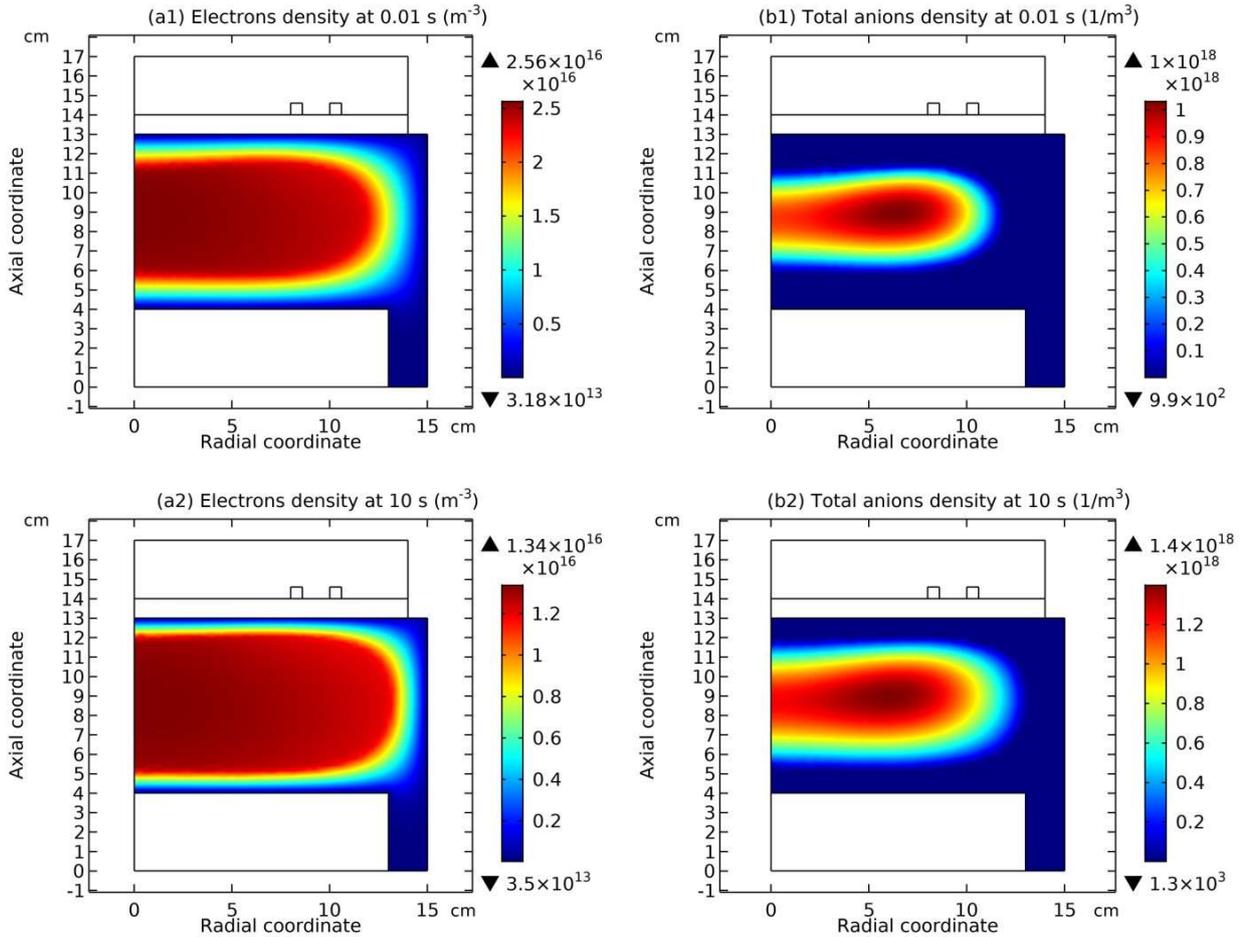

Figure 18. Simulated two-dimensional profile of electrons density at (a1) $0.01\,\text{s}$ and (a2) $10\,\text{s}$, and simulated two-dimensional profile of total anions density at (b1) $0.01\,\text{s}$ and (b2) $10\,\text{s}$, by the fluid model of Sec. (2.1). The discharge conditions of the Ar/SF$_6$ ICP are the same as in Figs. 1 and 2. The detailed temporal evolutions of the two-dimensional profiles of electrons density and total anions density given by the fluid simulation of Sec. (2.1) are shown in the video and supplementary files.

To ensure that the ionic and acoustic vibrations are physics, the convergence of simulation is studied in this section, and in this convergence the perfect harmony among the multiple processes contained in the discharge structure hierarchy, i.e., the negative anions chemical source, the parabola core, the double layer, the Boltzmann balance, the uniform profile of electrons density, the stratification, the ionic and acoustic vibrations, and even the wall sheath, is observed. Besides, the strong power of self-coagulation is detected, i.e., the larger is the volume of coagulated region, the denser is the coagulated body, which represents a geometric effect, actually, and is analogous to the astrophysical bodies.

In Figs. 18 and 19, the two-dimensional profiles of electrons density, total anions density, net source of anions, and net charge density at the two selected simulated times, $0.01\,\text{s}$ and $10\,\text{s}$, are given. In the process of convergence, the core region, represented by the uniform density profile of electrons, expands, as seen in Fig. 18(a1, a2), and in Fig. 18(b1, b2) the border of total anions density also expands and meanwhile the density of them increases, from $10^{18}\,\text{m}^{-3}$ to $1.4\times10^{18}\,\text{m}^{-3}$.

The increase of anions density in the expanding core is given by the expanding net source of anions in Fig. 19(a1, a2), which includes more regions of positive source. It is noted that the positive source of anions is in the periphery while the negative source of them is in the center. The larger is the radius, the larger is the volume surrounded, which represents a geometrical effect in the space. The self-coagulation is a dynamic balance of mass, i.e., input of anions by means of positive source in the peripherical region and coagulating them in the core by means of the negative source therein. So, the larger is the external region of positive source, the denser is the internal coagulated body therein formed. In Fig. 19(b1, b2), in the process of convergence, most of net charge density vibrations that are away from the border of core and halo are damped, while the strongest vibration at the border is left which plays the role of double layer that separates the core and halo. It is believed that the expansion of core is given by the self-coagulation process which is aimed at absorbing more discharge mass to form dense structure in the present hierarchy. Moreover, the acoustic vibration amplifies the expansion of core since the vibrations are destined to destroy the stable structure given by transport. Nevertheless, the discharge structure hierarchy sustains since the electropositive halo and sheath near the chamber border counteract the expansion of core, by means of the strong potential barrels of ambi-polar diffusion and sheath. This is thereby called the harmony of so many different processes in the hierarchy, i.e., the collective interactions of plasma.

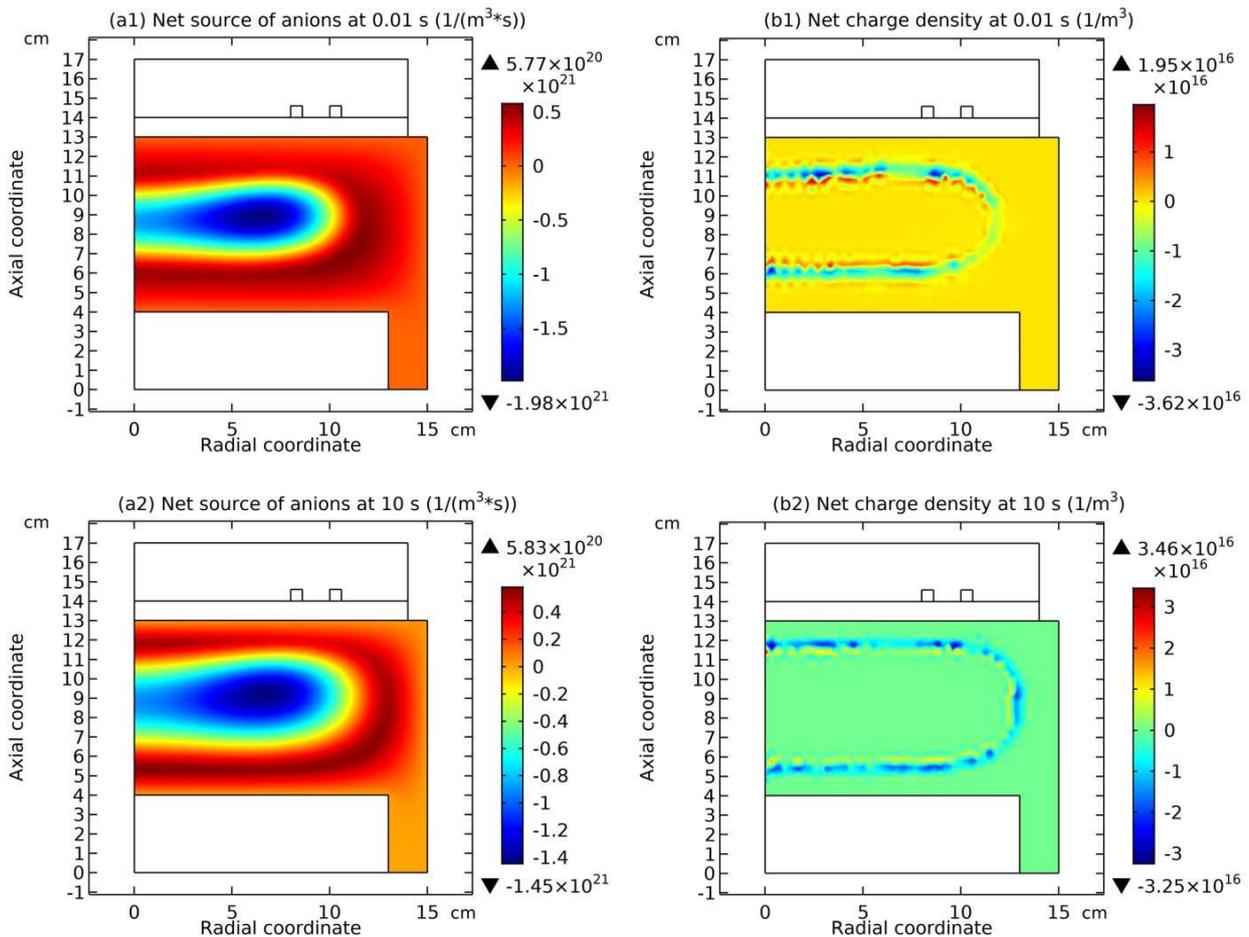

Figure 19. Simulated two-dimensional profile of net chemical source of anions at (a1) $0.01\,\text{s}$ and (a2) $10\,\text{s}$, and simulated two-dimensional profile of net charge density at (b1) $0.01\,\text{s}$ and (b2) $10\,\text{s}$, by the fluid model of Sec. (2.1). The discharge conditions of the Ar/SF$_6$ ICP are the same as in Figs. 1 and 2. The detailed temporal evolutions of the two-dimensional profiles of electrons density and total anions density given by the fluid

simulation of Sec. (2.1) are shown in the video and supplementary files.

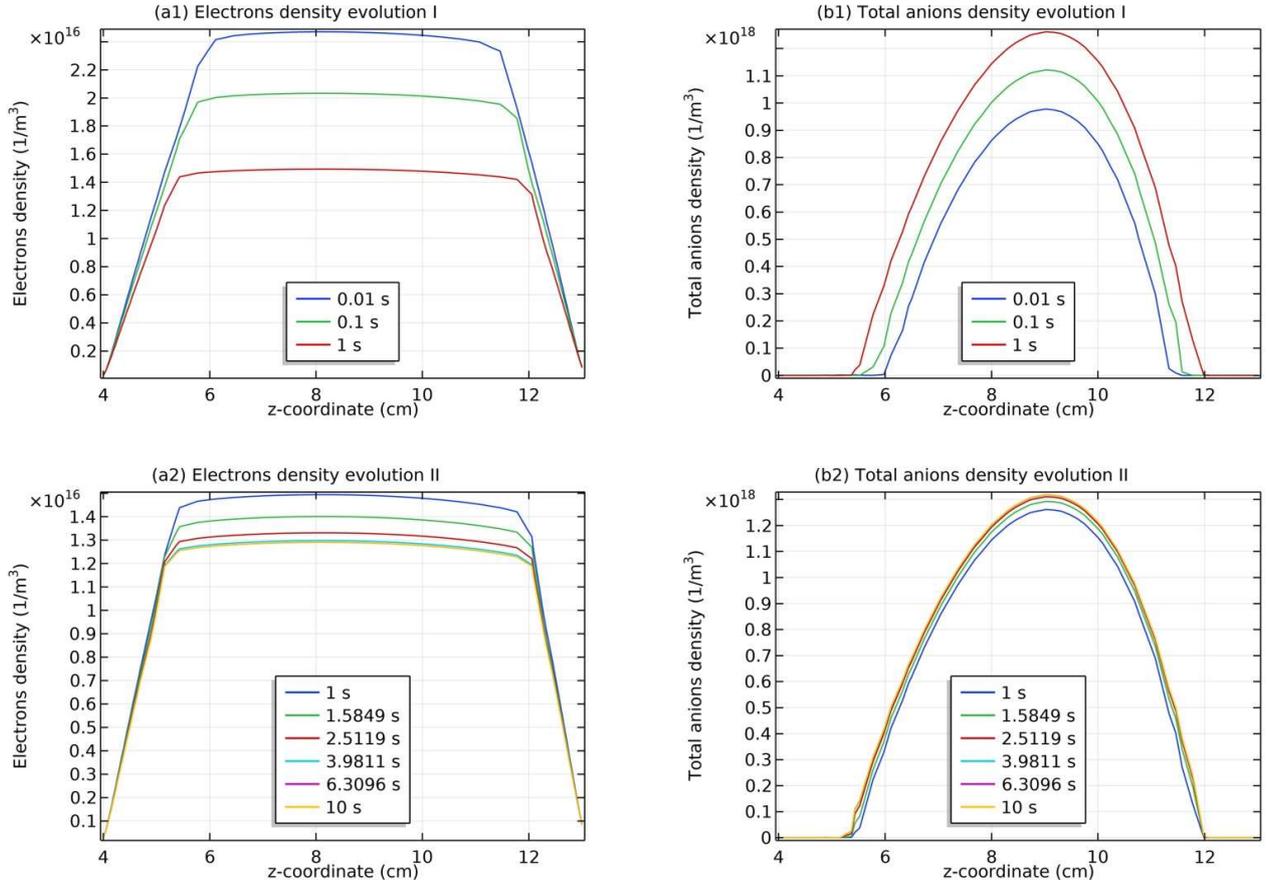

Figure 20. (a1) The first stage of temporal evolution of axial profile of electrons density simulated, i.e., evolution I and (a2) the second stage of temporal evolution of axial profile of it, i.e., evolution II, and (b1) the first stage of temporal evolution of axial profile of total anions density simulated, i.e., evolution I and (b2) the second stage of temporal evolution of axial profile of it, i.e., evolution II, given by the fluid model of Sec. (2.1) and at the radial position of 8.0 cm. The discharge conditions of the Ar/SF$_6$ ICP are the same as in Figs. 1 and 2. The detailed temporal evolutions of the two-dimensional profiles of electrons density and total anions density given by the fluid simulation of Sec. (2.1) are shown in the video and supplementary files.

In Figs. 20 and 21, the temporal evolutions of axial profiles of electrons density, total anions density, net source of anions, and net charge density in the fluid simulation of Sec. (2.1) are given. It is noted that the evolutions are divided into two parts, i.e., Evolution I with the time range of $0.01\,\text{s}$ to $1\,\text{s}$ and Evolution II with the time range of $1\,\text{s}$ to $10\,\text{s}$. The Evolution I gives the physics of plasma transport that forms the stable discharge structure hierarchy in Figs. 20(a1, b1) and 21(a1, b1) and the Evolution II gives the final convergence of these related quantities in Figs. 20(a2, b2) and 21(a2, b2). In the Evolution I part of Figs. 20(a1, b1), the electrons density decreases while the total anions density increases in the expansion of core, which reveals the further transfer of negative charge carrier of this plasma source from the electrons to anions. It is noted the electrons density decrease herein supplements the anions positive source growth from $5.77 \times 10^{20}$ m$^{-3}$s$^{-1}$ to $5.83 \times 10^{20}$ m$^{-3}$s$^{-1}$ in Fig. 19(a1, a2), to complete the above transfer of negative charge carrier. In Fig. 21(a1, a2), the magnitude of negative anions chemical source decreases and finally converges, indicating when the volume of self-coagulation region is expanded, less negative chemical source is enough to sustain the coagulation. Last, in Fig. 21(b1, b2), the ionic and acoustic vibrations away from the border are smoothed by the diffusion of ionic species, along with the expansion of core, and then they concisely converge. The tiny instability marked in Fig. 21(b2) will be discussed in Sec. (3.5).

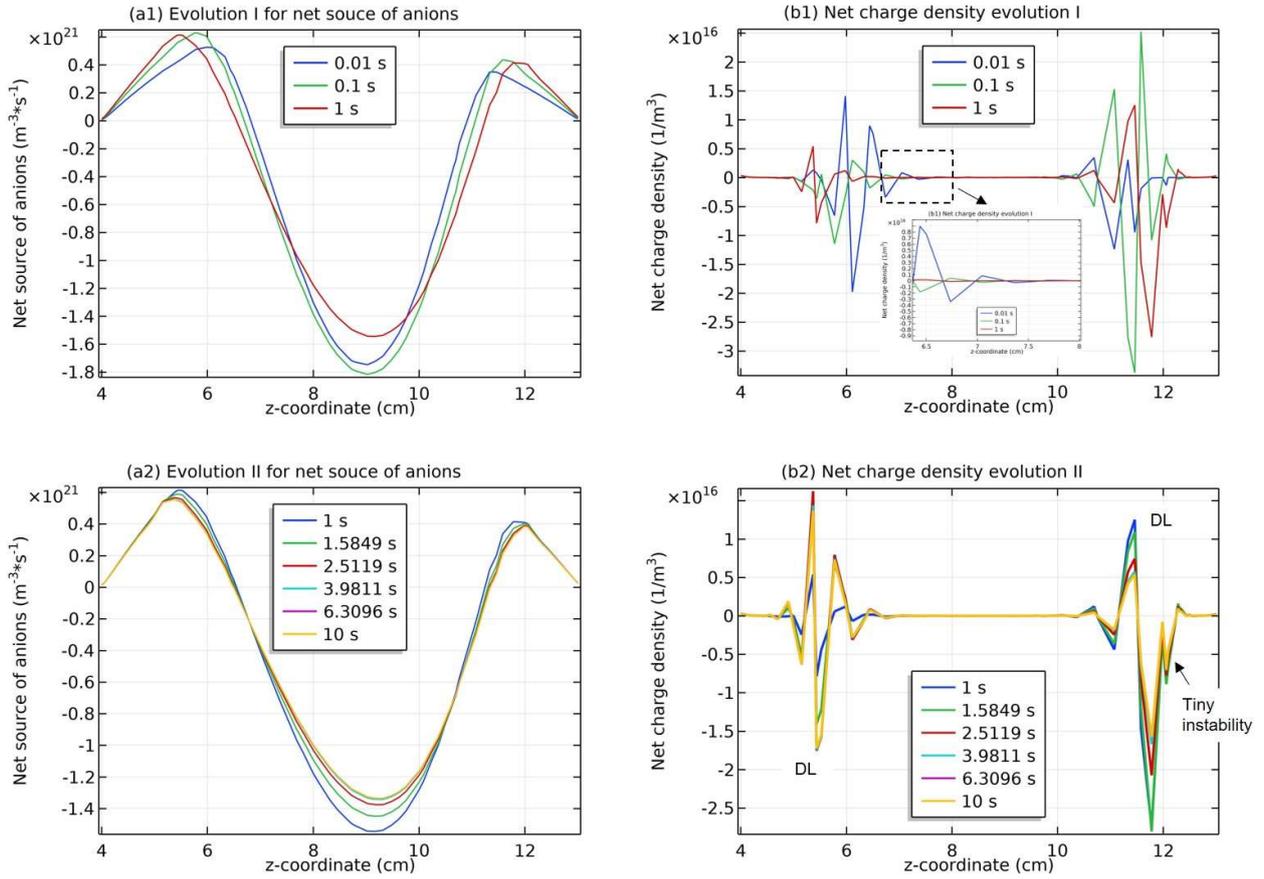

Figure 21. (a1) The first stage of temporal evolution of axial profile of net source of anions simulated, i.e., evolution I and (a2) the second stage of temporal evolution of axial profile of it, i.e., evolution II, and (b1) the first stage of temporal evolution of axial profile of net charge density simulated, i.e., evolution I and (b2) the second stage of temporal evolution of axial profile of it, i.e., evolution II, given by the fluid model of Sec. (2.1) and at the radial position of 8.0 cm. The discharge conditions of the Ar/$SF_6$ ICP are the same as in Figs. 1 and 2. The inserted sub-figure in Panel (b1) clearly shows the damping process of net charge vibration with time in the core. The detailed temporal evolutions of the two-dimensional profiles of electrons density and total anions density given by the fluid simulation of Sec. (2.1) are shown in the video and supplementary files. The tiny instability marked in Panel (b2) will be discussed in Sec. (3.5).

## (3.5) An overview of hierarchy of high-pressure case, i.e., self-coagulation, ellipse profile, blue sheath and instabilities

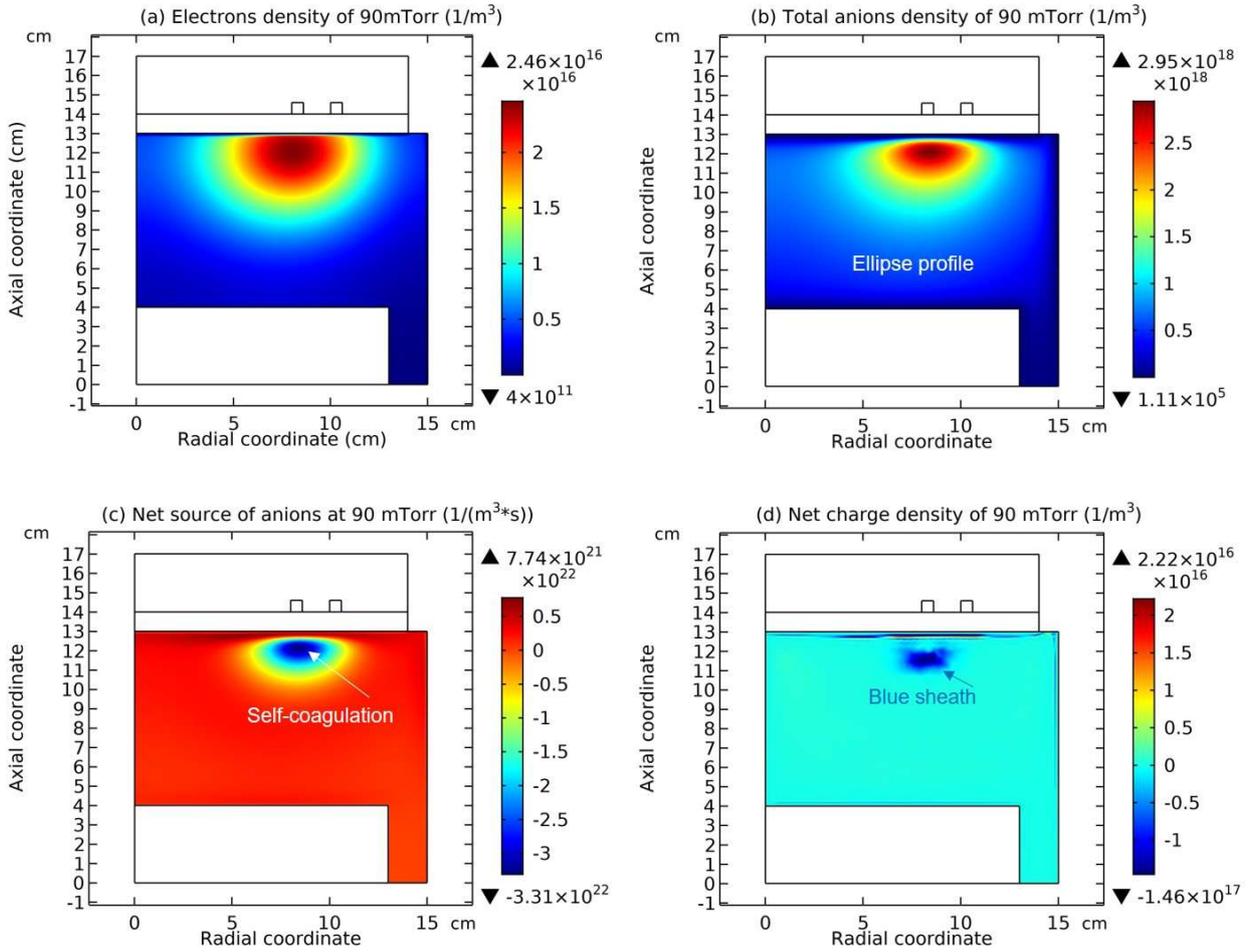

Figure 22. Simulated two-dimensional profiles of (a) electrons density, (b) total anions density, (c) net source of anions, and (d) net charge density at 90 mTorr, given by the fluid model of Sec. (2.1). The other discharge conditions of the Ar/$SF_6$ ICP are 300W and 10% $SF_6$ content, and the simulated time is 0.1 s.

Fig. 22 shows the discharge structure hierarchy of Ar/$SF_6$ ICP simulated by the fluid model of Sec. (2.1) at the high pressure, 90mTorr, and in this figure the two-dimensional profiles of the electrons density, the total anions density, the net source of anions, and the net charge density are sequentially presented. The stratification disappears at the high pressure because the transport becomes slow and the strong electrons consumption scheme, attachments, blocks the buildup of a spatially symmetric ambi-polar potential, and hence the electrons density profile is localized under the coil as shown in Fig. 22(a). For the total anions density of Fig. 22(b), except for the coagulated semi-circle under the coil, the ellipse profile of them dominates in the discharging space, which is predicted by another analytical theory given in Ref. [26]. As for the net source of anions in Fig. 22(c), the predominant negative source of anions under the coil implies the self-coagulation process happening therein, and for the net charge density of Fig. 22(d), the non-neutral region under the coil indicates the appearance of an electronegative sheath, i.e., blue sheath, instead of the double layer anymore. In Fig. 23, the temporal evolution of net charge density axial profile under the coil is shown, both locally in an axial range of 10.45 and 12.5 cm and globally in a big axial range of 9 and 12.5 cm. It is seen from Figs. 22(d) and 23(a) that besides for the blue sheath, the instabilities

**33** / 46

appear, and actually, the blue sheath of Fig. 22(d) corresponds the strongest instability, i.e., the third one in Fig. 23(a). In addition, it is still noted that the instabilities are only predominant in the self-coagulation region (see Fig. 22(c) for reference), i.e., the local axial region of 10.45 and 12.5 cm as shown in Fig. 23(a), and away from the coagulation region, i.e., smaller than the 10.45 cm, the instabilities are not grown, as shown in Fig. 23(b). It is analyzed that the instabilities in the coagulation region belong to the type of ionization instability [34-41] since in this region the chemical source of cations is positive, as seen from Fig. 24(b, c) where the two-dimensional profile of net source of cations and the related axial profile of it under the coil are shown. Specially, in Fig. 24(c), the border between the negative (dominance of recombinations) and positive (dominance of ionizations) chemical sources of cations is marked, which is more or less in accord to the border of insignificant and significant instabilities in Fig. 23(b).

    It is concluded from the discharge structure hierarchies of the high and low pressures that the self-coagulation plays the role of turning the gaseous discharging plasmas into a collisionless plasma, in which the quasi-astrophysical plasma behaviors occur, i.e., the acoustic type of double layer at the low pressure and the instability at the high pressure when the chemical ionizations play the role of free energy that is added into such a physically elastic-collisionless plasma [34]. It is stressed that the occurrence of instability at the high pressure is not easy since it requires the collaboration between the negative source of *anions* that creates the collisionless plasma through the self-coagulation and the positive source of *cations* that provides the free energy through the chemical kinetics. Coincidently, these two required conditions are satisfied in the same region, i.e., under the coil as displayed in the simulation at the high pressure. At low pressure, although the stratified plasma is also collisionless by means of self-coagulation, the position of the interface of core and halo is far away from the ionization peak of cations, which is also under the coil as simulated. Hence, the instability is not occurred due to the lack of free energy and the double layer evolves therein through the ionic and acoustic vibrations, at least in an axial range of 4 and 7 cm away from the coil as shown in the left side of Fig. 21(b2). It is seen that the similar ionization instability appears in another axial range of 10 and 12.5 cm close to the coil as shown in the right side of Fig. 21(b2), but the magnitude of this instability is weak, as marked in the figure.

    Besides, the total cations density simulated at 90 mTorr in Fig. 24(a) holds the similar profile as the anions in Fig. 22(b). Regarding the novel profiles of net chemical sources of anions and cations in Figs. 22(c) and 24(b), i.e., both the superpositions of negative and positive sources, the self-coagulations of ions at the high pressure are mono-polar, which are more determined by the chemical kinetics, different with the ambi-polar one at the low pressure that depends on the electrical interaction as described in Sec. (3.1). More details about the discharge structure hierarchy of high pressure will be given in future articles.

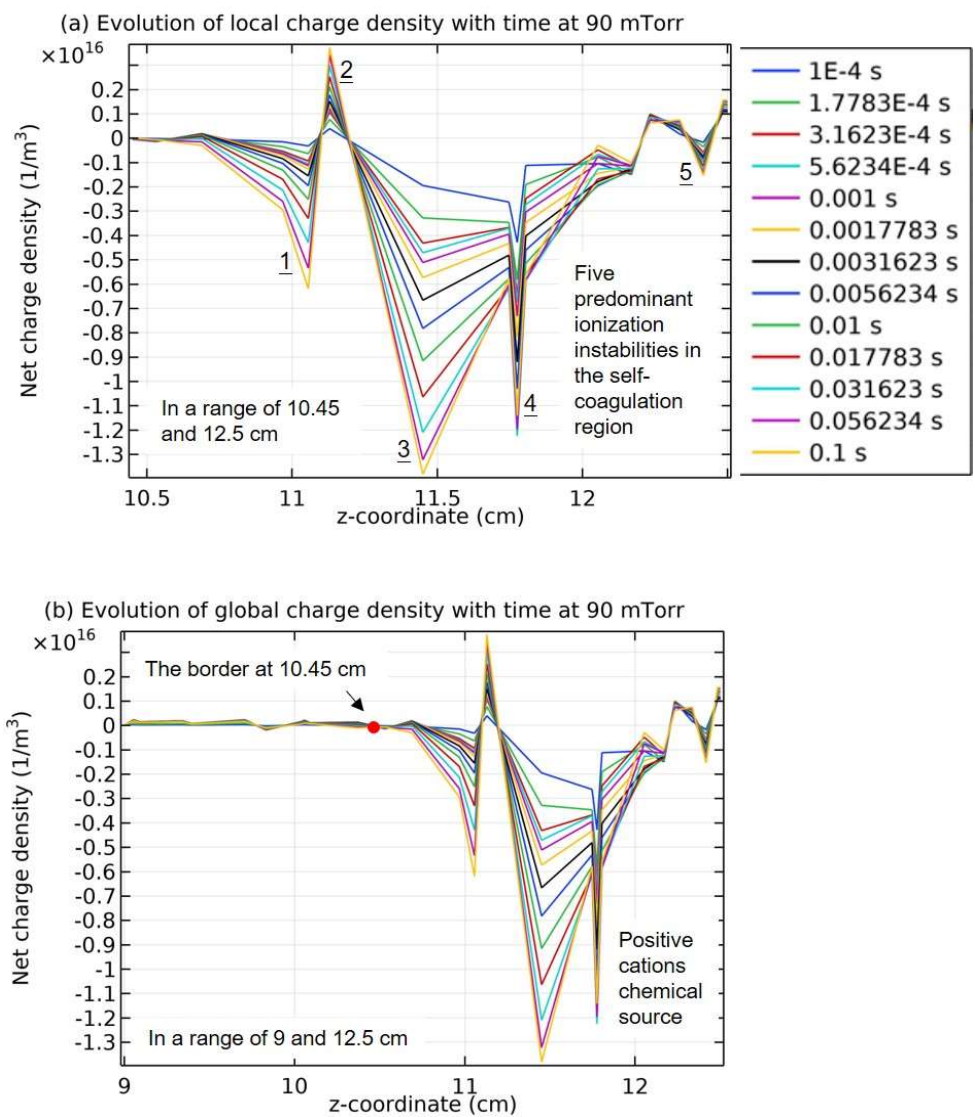

Figure 23. Evolution of axial profile of *local* charge density with time, in an axial range of 10.45 and 12.5 cm, and (b) evolution of axial profile of *global* charge density with time, in a big axial range of 9 and 12.5 cm, both given by the fluid simulation of Sec. (2.1) at the high pressure, 90 mTorr, and under the coil. Hence, the radial position of these axial charge density profiles is 7.9 cm. The temporal evolutions of local and global net charge density axial profiles share the same legend given in Panel (a). The other discharge conditions of the Ar/$SF_6$ ICP are 300W and 10% $SF_6$ content, and the simulated time is 0.1 s. It is noted that the instabilities are predominant in the self-coagulation region, as shown in Panel (a), and away from this region the instabilities are not grown, as shown in Panel (b). The instabilities in Panel (a) belong to the type of ionization instability since in the coagulation region, the cations chemical source is positive, as shown in Fig. 24(b, c).

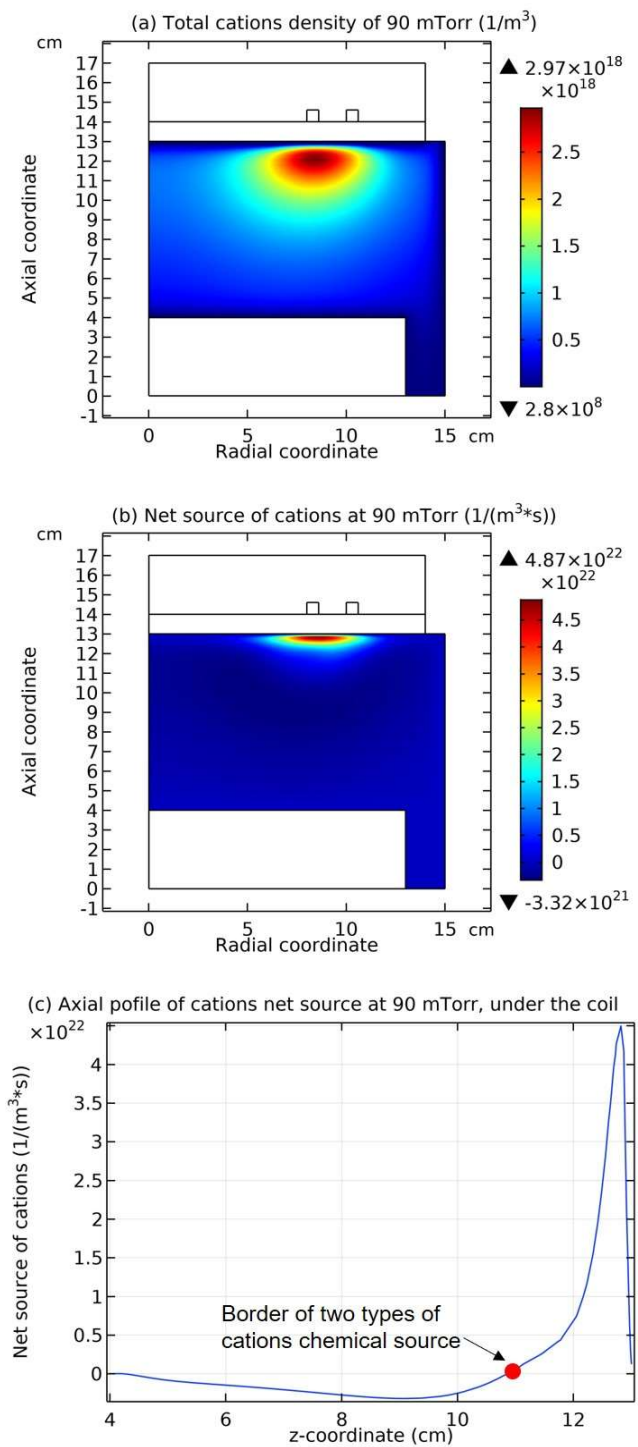

Figure 24. Two-dimensional profiles of (a) total cations density and (b) net source of cations, and (c) axial profile of cations net source with the radial position, 7.9 cm, given by the fluid simulation of Sec. (2.1) at 90 mTorr. The other discharge conditions of the Ar/$SF_6$ ICP are 300W and 10% $SF_6$ content, and the simulated time is 0.1 s. In Panel (c), the border between the negative and positive chemical sources of cations is marked, at the two sides of which the ionic recombinations and the ionizations dominate, respectively.

## (3.6) Process of turning the discharging plasma into collisionless plasma of anions self-coagulation

$$m_- \frac{\partial n_- \vec{u}_-}{\partial t} + m_- \nabla \cdot (n_- \vec{u}_- \vec{u}_-) = -kT_- \nabla n_- - en_- \vec{E}_{real} - n_- \vec{u}_- n_n v_{-n}, \tag{47}$$

$$m_- \frac{\partial n_- \vec{u}_-}{\partial t} + m_- \nabla \cdot (n_- \vec{u}_- \vec{u}_-) \sim 0,\ \mu_- = \frac{e}{n_n v_{-n}},\ D_- = \frac{kT_-}{n_n v_{-n}},\ \vec{\Gamma}_- = n_- \vec{u}_-, \tag{48}$$

$$\vec{\Gamma}_- = -D_- \nabla n_- - \mu_- n_- \vec{E}_{real}, \tag{49}$$

$$\Gamma_- \sim 0,\ -D_- \nabla n_- - \mu_- n_- \vec{E}_{real} = 0, \tag{50}$$

$$n_- \vec{u}_- n_n v_{-n} = -kT_- \nabla n_- - en_- \vec{E}_{real}, \tag{51}$$

$$-kT_- \nabla n_- - en_- \vec{E}_{real} \sim 0,\ n_- \vec{u}_- n_n v_{-n} \sim 0. \tag{52}$$

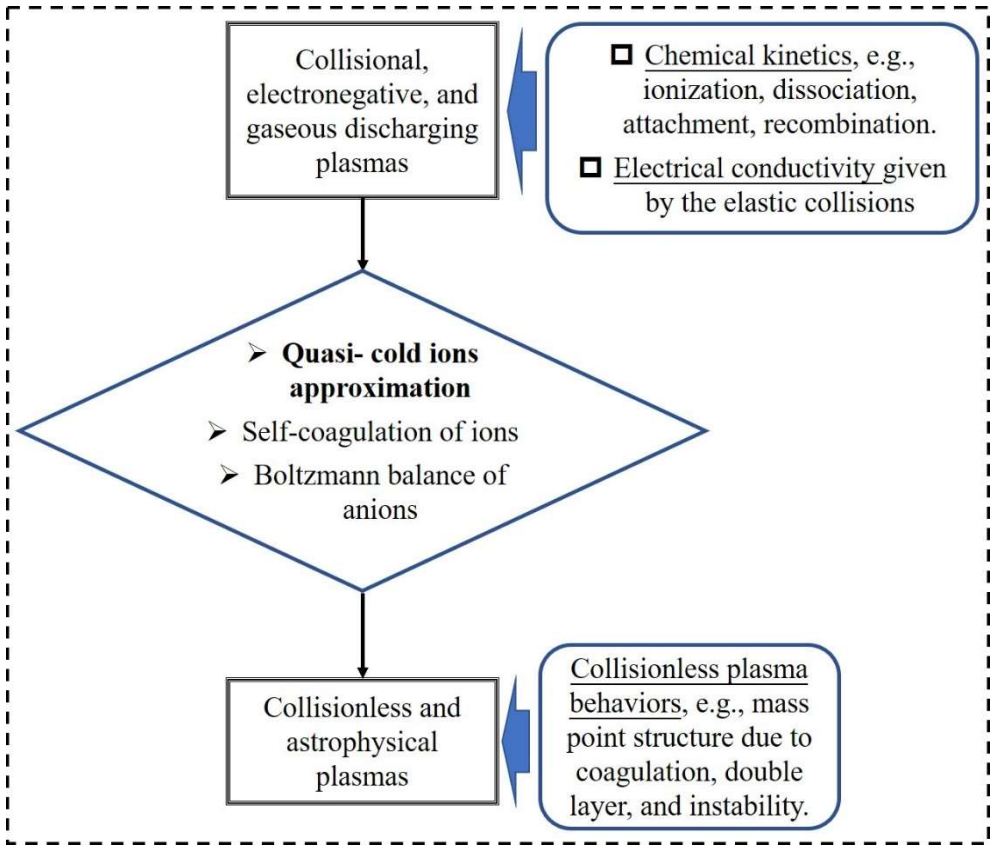

Figure 25. Flowchart on the process of turning collisional, electronegative, and gaseous discharging plasmas into collisionless and astrophysical plasmas by the self-coagulation that only occurs when the ions are approximated to be cold (see next), and the chemical kinetics of collisional plasmas and its electrical conductivity due to the elastic collisions, and the behaviors of collisionless plasmas, as listed.

In this section, it is interpreted how the self-coagulation turning the discharging plasma into the collisionless plasma via the Boltzmann balance. In Eq. (47), the complete momentum equation of anions is given, which sequentially describes the inertial, advection, diffusion, drift and collisional terms. In Eq. (48), when the inertial and advection terms are neglected and the transport coefficients, mobility and diffusion, and the flux of anions are introduced, the complete momentum equation can be simplified into the approximated drift and diffusion flux in Eq. (49). As mentioned in Sec. (3.2), the anions Boltzmann balance is triggered by the self-coagulation. So, in Eq. (50), the flux of anions is assumed to be zero and then the counteracted diffusion and drift fluxes generate the anions Boltzmann balance. In Eq. (51), the approximated drift and diffusion flux is rewritten and in Eq. (52), it is seen that the collisional term automatically disappears when the Boltzmann balance is satisfied in such a newly written flux. In a word, when the self-coagulation occurs, it induces the Boltzmann balance and the transfer of gaseous discharging plasma into collisionless plasma. In Fig. 25, the flowchart of such a process is illustrated. It is stressed in this figure that the self-coagulation occurs at the quasi- cold ions approximation, and when the ions are thermalized by the Ohm heating mechanism based on the electrical conductivity the hierarchy is replaced by a simple discharge structure (see next). Besides, it is noted that in the collisionless plasma generated by the self-coagulation, the elastic collisional term corresponding to the electrical conductivity is insignificant in the plasma transport and heating mechanism, not the inelastic collisions, such as the ionization, dissociation, attachment and recombination as listed in Fig. 25, which represent the chemical kinetics of discharging and collisionless plasmas and thereby sustain them. More analysis on distinguishing the elastic and inelastic collisions in the quasi- collisionless plasma will be given in Sec. (3.8).

## (3.7) Simple discharge structure when the ions are thermalized in the fluid model

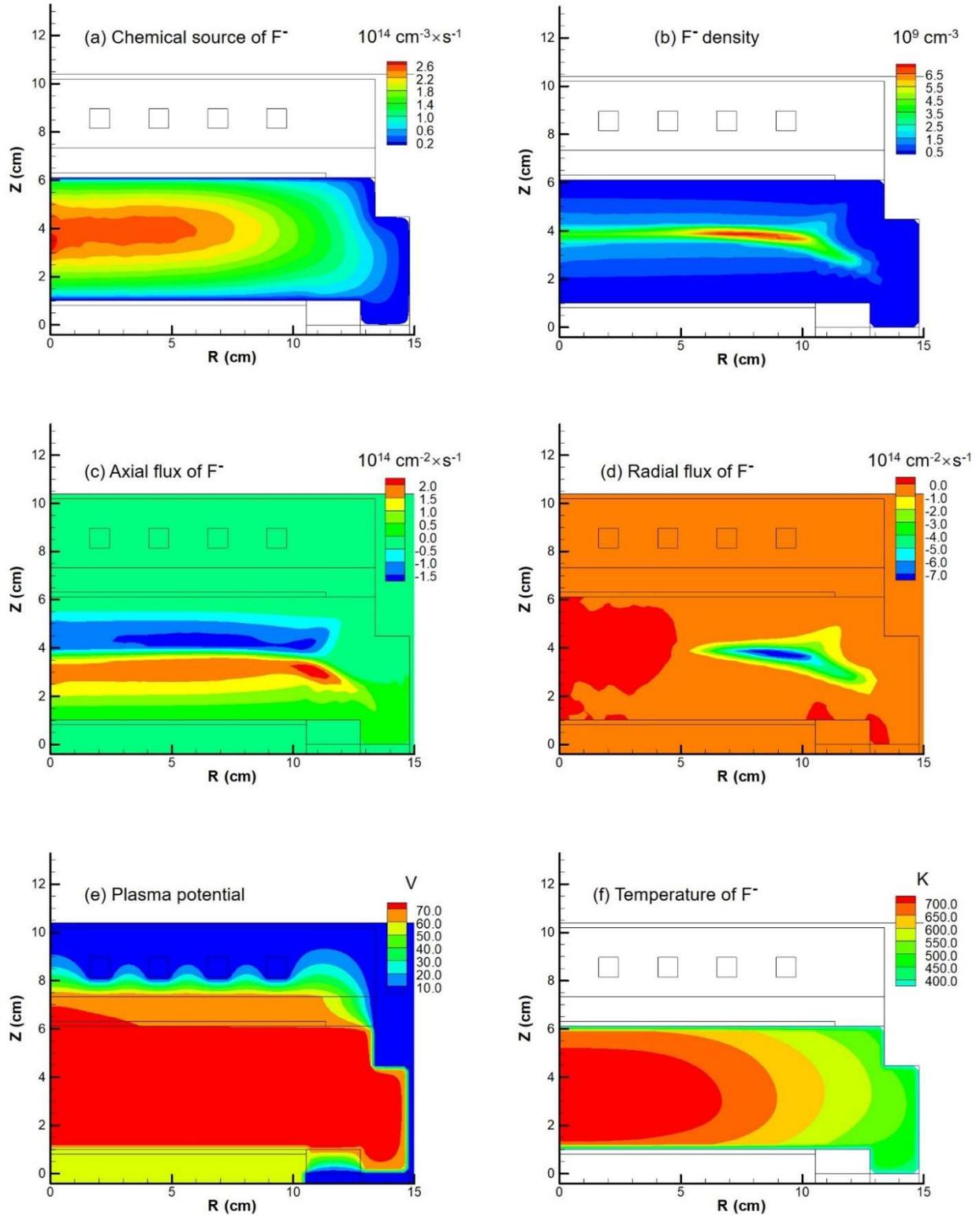

Figure 26. Simulated two-dimensional profiles of (a) chemical source of $F^-$, (b) $F^-$ density, (c) axial flux of $F^-$, (d) radial flux of $F^-$, (e) plasma potential, and (f) temperature of $F^-$ in a mixed Ar/$CF_4$ ICP source, given by the quasi-fluid simulation of HPEM code that is described in Refs. [43, 44] and in the acknowledgement. The discharge conditions of this ICP source are 500W, 10mTorr and 50% $CF_4$ content. Besides, the substrate is biased and more information about this bias can be found in Ref. [44].

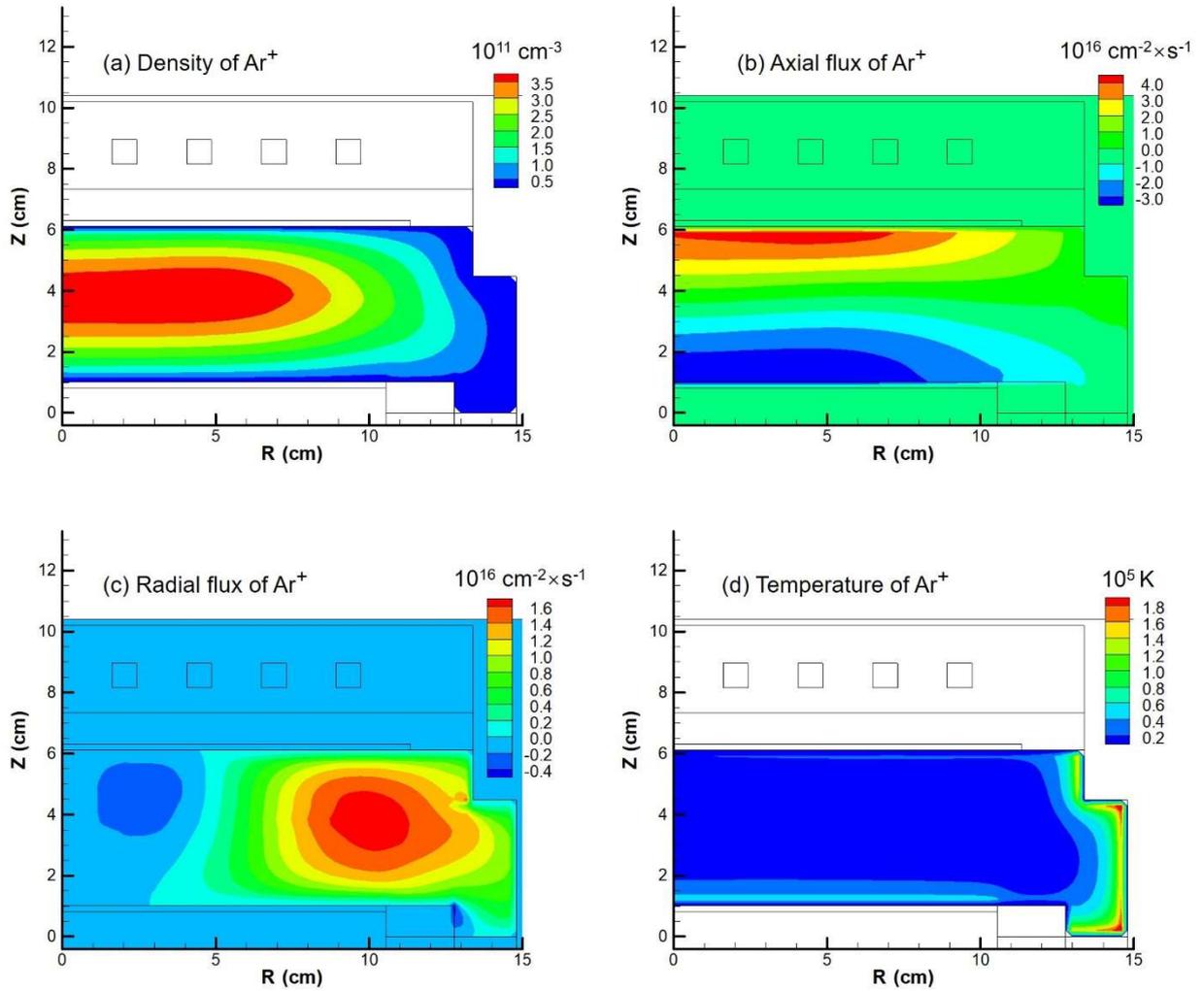

Figure 27. Simulated two-dimensional profiles of (a) $Ar^+$ density, (b) axial flux of $Ar^+$, (c) radial flux of $Ar^+$, and (d) temperature of $Ar^+$ in the mixed $Ar/CF_4$ ICP source, given by the quasi-fluid simulation of HPEM code that is described both in Refs. [43, 44] and in the acknowledgement. The discharge conditions of this source are 500W, 10mTorr and 50% $CF_4$ content. Besides, the substrate is biased and more information about this bias can be found in Ref. [44].

    In this section, it is shown that when the ions are thermalized by the Ohm heating mechanism that relies on the electrical conductivity of elastic collisions between the ions and neutrals, the discharge structure hierarchy disappears and the structures of electronegative ICPs are more like the electropositive ones. In Fig. 26, the chemical source of $F^-$ species, the density of it, the axial and radial fluxes of it, the plasma potential, and the temperature of it in an $Ar/CF_4$ ICP source are presented, which are calculated by the HPEM code [43] that includes the heavy species thermal energy equations, different with the fluid model set in Sec. (2.1). In Fig. 27, the density of $Ar^+$, the axial and radial fluxes of it, and the temperature of it in the $Ar/CF_4$ ICP are given, also calculated by the HEPM code. The discharge conditions can be found in the captions of Figs. 26 and 27. The ions fluid equation set in the HPEM code are given in the Sec. (3.8), and it includes the ions energy equation. It is seen in Fig. 26(a) that the chemical source of $F^-$ anions is positive and hence the self-coagulation is not formed, neither of the discharge structure hierarchy. The density peak of $F^-$ in

Fig. 26(b) is given by the pushing and squeezing roles of potential barrel in Fig. 26(e), which hence forms the axial and radial fluxes of F⁻ in Figs. 26(c, d). So, this is a physical coagulation by means of the external force, not a self-coagulation that relies on the internal and chemical force, i.e., negative chemical source. In Fig. 26(f), the temperature of F⁻ is low in the peripheric regions and increases with the distance to the discharge center, due to the Ohm heating scheme of ambi-polar potential barrel in Fig. 26(e). The central temperature of F⁻ calculated by the ions energy equation of HPEM code is around 700K, rather higher than the room temperature of anions designed in the fluid model of COMSOL software, 300K, in Sec. (2.1). In Fig. 27, the picture of cations, e.g., Ar⁺, in such a simple discharge structure is presented. The symmetric profile of Ar⁺ density is shown in Fig. 27(a), and in Fig. 27(b, c) it is seen that the axial and radial fluxes of Ar⁺ are contradictory to the anion, F⁻, i.e., the ambi-polar potential pulls the Ar⁺ cations outwards while pushes the F⁻ anions inwards. Since the Ar⁺ cations are in general transported out to the chamber wall, the temperature of them peaks therein, as shown in Fig. 27(d). It is seen that the peaked temperature of cations, Ar⁺, is rather higher than the anions, F⁻, i.e., $1.9 \times 10^5$ K versus 700 K. This is because the cations, Ar⁺, experience the strong Ohm heating scheme in the sheath during their ways to the chamber wall, especially in the corner of chamber. It is noted that the simple discharge structure of electronegative ICP calculated by the HPEM code is real, since the Ohm heating scheme of ions indeed exists and hence the ions are thermalized. However, as seen next, the simple discharge structure with the thermal ions probably can be transferred to the hierarchy of structure by decreasing the gas flow rate, which is designed to decrease the Ohm heating scheme, to quench the thermal ions and hence to trigger the self-coagulation and Boltzmann balance possibly.

## (3.8) Construction of fluid model for nuclear fusion based on self-coagulation, i.e., without inertial or magnetic confinement, and the related experimental design

$$\frac{\partial n_i}{\partial t} + \nabla \cdot (n_i \vec{u}_i) = R_i, \tag{53}$$

$$\frac{\partial n_i m_i \vec{u}_i}{\partial t} + \nabla \cdot (n_i m_i \vec{u}_i \vec{u}_i) = -\nabla (k n_i T_i) + e n_i \vec{E}_s - \vec{M}_i, \tag{54}$$

$$\frac{\partial (n_i c_i T_i)}{\partial t} + \nabla \cdot (n_i \vec{u}_i c_i T_i) = \nabla \cdot (\kappa \nabla T_i) - P_i \nabla \cdot \vec{u}_i + \frac{n_i q_i^2}{m_i \nu_{in}} E_s^2 + E_i, \tag{55}$$

$$\vec{M}_i = \sum_n \frac{m_n}{m_i + m_n} n_i n_n (\vec{u}_i - \vec{u}_n) \nu_{in}, \tag{56}$$

$$E_i = \sum_n 3 \frac{m_n}{m_i + m_n} n_i n_n k (T_n - T_i) \nu_{in} - \sum_n 3 n_i n_n R_{in} k T_i. \tag{57}$$

In Eqs. (53-57), the complete fluid equation set for the ions in the HPEM code are given. Sequentially, the continuity equation in Eq. (53), the momentum balance equation in Eq. (54), the thermal energy balance equation in Eq. (55), and the expressions for the collisional damping term, $\vec{M}_i$, in the momentum equation in Eq. (56) and for the thermal energy loss term due to non-thermal equilibrium and charge exchange collision, $E_i$, in the energy equation in Eq. (57) are presented. Specially, in the energy equation of Eq. (55), the compresible heating term in blue color, $-P_i \nabla \cdot \vec{u}_i$, and the Ohm heating term in red color, $\frac{n_i q_i^2}{m_i \nu_{in}} E_s^2$, are given. It is noted that the Ohm heating mechanism is the predominant factor in the gaseous discharging glow plasma, i.e., low-temperature plasma, while the compressible heating mechanism is the predominant factor in the nuclear fusion plasma, i.e., high-temperature plasma. The collision is not related in the compressible heating term, while the Ohm heating term relies on the elastic collision frequency between ions and neutrals, $\nu_{in}$. More details about the ions equation set can be found in Refs. [43, 45].

From the Secs. (3.1-3.5), it is known that the when the ions are cold in the electronegative ICP sources, the self-coagulation and the related complex discharge structure hierarchy are formed, as given by the COMSOL software. Nevertheless, when the ions energy equation is used, the simple and quasi- electropositive discharge structure is formed by the HPEM code, as shown in Sec. (3.7). Based on the analysis of energy equation, Eq. (55), it is proposed that the nuclear fusion plasma possibly can be formed in the laboratory plasmas without the magnetic and inertial constrictions. It is designed that the chamber with movable wall is built. Firstly, the gaseous glow discharging plasma is excited. As mentioned in Sec. (3.7), in the glow discharging plasma the ions are thermalized by the Ohm heating mechanism, and the discharge structure is simple and has no hierarchy. Then, the gas flow rate is reduced to decrease the collision frequency and then the Ohm heating scheme. It is hoped this operation can reduce the temperatures of ions and hence form the discharge structure hierarchy. After the hierarchy is formed, the chamber wall is expanded and the discharge volume increases, which based on the ambi-polar self-coagulation, perhaps forms higher ions densities, by means of the physics analyzed in the Fig. 20(b1) of Sec. (3.4), i.e., the larger is the volume of coagulated body, the higher is the density of it. Further relying on the compressible heating mechanism, the temperature and density of self-coagulated quasi- astrophysical body can be both high enough to realize the nuclear fusion of discharging mass. The essence of such a self-nuclear fusion is the transfer of Ohm heating mechanism to the compressible heating mechanism [46] by means of decreasing the gas flow rate. It is estimated to be successful since the chemical reaction kinetics rely on the inelastic collisions while the Ohm heating mechanism relies on the elastic collisions. It is known that the inelastic collision frequency between the ions and neutrals is several orders lower than the elastic collision, and hence the chemical kinetics are more insensitive to the variation of gas flow rate than the Ohm heating scheme. It is hoped that the gaseous glow discharging plasma will not extinguish when decreasing the gas rate, but turn into the arc plasma at first that has no the ionic Ohm heating and compressible heating mechanisms [47-49] and then into the fusion plasma at absorbing more and more mass by means of self-coagulation and compressible heating scheme [46]. It is noted that in the novel fusion process, the self-coagulation plays the role of gravitational force in the formation of astrophysical body.

# VI. Conclusion and further remarks
## (a) Conclusion

When the ions are thermalized by the Ohm heating scheme, the anions are insignificant in forming the main-body discharge structure. The discharge structure is mainly sustained by the electrons and cations, and the general ambi-polar diffusion potential barrel is formed and the anions are pressed into the discharge interior by it, hence less participating in the discharge process. This simple discharge structure is hence called as the quasi- electropositive one, and the plasma is collisional and can be described by the HPEM code.

When the ions are quenched, the complex discharge structure hierarchy is formed, which consists of the self-coagulation, the Boltzmann balance, the stratification, the parabola core, the double layer, and the instability, etc. This plasma is collisionless and can be described by the COMSOL software. It is of significance in both the astrophysics and the nuclear fusion physics. As estimated, at certain circumstance, the collisional plasma can be transferred into the collisionless plasma by means of replacing Ohm heating scheme with the compressible heating scheme, which suggests to realize the self-fusion in the laboratory that does not need the constrictions of magnetic field and inertial term, relying on the self-coagulation and related discharge structure hierarchy.

In the Comsol software simulation part of the article, the quasi- cold ions approximation suits to the ions that are in the room temperature, not the zero temperature. So, the diffusion that relies on the room temperature of ions is included, which smooths the ionic and acoustic vibrations away from the double layer. Besides, it is noted that the different roles of ions temperature and their heating mechanisms are the clue of the article outline, i.e., (i) the concise cold ions approximation used in the analytics and semi- analytics of the stratification, the core parabola, and the double layer, (ii) the quasi- cold ions approximation used in the fluid plasma module of COMSOL software, (iii) the realistic ions thermalization given by the Ohm heating scheme as simulated by the HPEM code, and (iv) the predictive and extreme ions thermalization of fusion given by the experimental design and the transfer of heating schem from the Ohm to compression.

The present article is focused on the discharge structure hierarchy at the low pressure of electronegative ICPs, 10mTorr. It is seen that when the pressure is increased to 90mTorr, the self-coagulation, and the Boltzmann balance of anions at least in the coagulated region, still exist, but the stratification disappears. Besides, the main profile characteristic changes from the parabola to the ellipse, predicted by the related analytical theories. The double layer is not occurred when the stratification disappears at high pressure, but the blue sheath and instability appear in the coagulated region, implying the advantage of self-coagulation in turning the high-pressure discharging plasma in which the collisions are significant to the collisionless plasma.

The detailed pictures of double layer, i.e., the double-valued property of plasma edge flux and potential and their relation, and the transferred dipole and capacitor models of double layer, are well predicted by both the simulation and analytics. The self-coagulation is the main factor of the hierarchy and it offers new insights on re-recognizing the origin of anions Boltzmann balance. The plasma collective interaction that leads to the harmony of so many sub-processes in the hierarchy is presented. The cooperation between the numerics and analytics is necessary for understanding the profiles of complex systems, e.g., electronegative discharging plasma, and the related physics.

## (b) Further remarks

It is noted that the inertial and advection terms are omitted in the momentum equation of COMSOL software, as illustrated by the Eq. (3) of Sec. (2.1). As estimated, the addition of these two terms accelerates the self-coagulation process, by means of providing better pre-condition for it, i.e., accumulating anions faster to give higher anions density and then form negative chemical source. The mesh size of finite element numerical method in the COMSOL software is decreased from *finer* to *extra fine*, and all the characteristics of discharge structure hierarchy in such a plasma source, e.g., self-coagulation, stratification, double layer, parabola core, and even ionic and acoustic vibrations, are kept (see the Figs. S5 and S6 in the supplementary material). The hierarchy is thus numerically stable. As mentioned before in Sec. (3.1), the self-coagulation in this article is ambi-polar, which is essentially the electrical interaction, like the ambi-polar diffusion. When increasing the pressure, the self-coagulations of ions are mono-polar. Concretely, the anions hold the central negative source while the peripherical regions provide anions by means of positive source, and the cations hold central positive source while the peripherical regions consume the generated cations by means of negative source. As seen, the two chemical schemes both generate coagulated structure. Moreover, besides for the ions, the electrons and even the neutrals have the self-coagulation process as well once the required condition is satisfied. The different self-coagulation types will be reported in future articles.

**Acknowledgements**

The data of Sec. (3.8) in the article are given by the HPEM simulation that was done by Shuxia Zhao when she was post-doc in the University of Antwerp, around ten years ago. During that time this simulation was to investigate the effect of gas ratio on the etch rate and radial uniformity in the Ar/$CF_4$ ICP source, as illustrated in Ref. [44]. Coincidently, the chemical source of representative anion and cation, the densities and fluxes of them, the plasma potential, and the temperatures of them given by that HPEM simulation can be used in the present article, to illustrate that the complex discharge structure hierarchy disappears and the simple discharge structure appears when the ions are thermalized by the Ohm heating scheme. The full name of HPEM is hybrid plasma equipment model and it is built by the Prof. Kushner and his team. Different with the fluid model set in Sec. (2.1) of the present article, the electron energy is not solved in the fluid kinetic module of HPEM code, but given by an electron Monte Carlo module of it used to account for the nonlocal effect of electrons. Hence, the HPEM code is called quasi- fluid model, as defined in the present article. Besides, the heavy species energy fluid equations are included in the fluid kinetic module of HPEM code and so the ions and neutrals can be thermalized by the Ohm heating scheme and charge-exchange collisions, respectively, which differs with the quais- cold ions and neutrals approximation designed in the fluid model of Sec. (2.1) in the present article (i.e., the general feature of plasma modules in the COMSOL software). It is believed that the ionic thermalization, not electron nonlocal effect, leads to the transfer of complex discharge hierarchy of electronegative ICP source to the simple structure, as proved by the early analytical works of Refs. [19, 20, 22].

Due to the strong support of the previous HPEM simulation in the University of Antwerp, the authors of present article are grateful to the teams of Prof. Bogaerts (the supervisor of Shuxia Zhao when she was in the University of Antwerp) and Prof. Kushner.